\documentclass[prx,onecolumn,floatfix,longbibliography,notitlepage, nofootinbib]{revtex4-1}

\usepackage[letterpaper,top=2cm,bottom=2cm,left=3cm,right=3cm,marginparwidth=1.75cm]{geometry}

\usepackage{graphicx, color, graphpap}
\usepackage{enumitem}
\usepackage{amssymb}
\usepackage{amsthm}
\usepackage{dsfont}
\usepackage{mathtools}
\usepackage[dvipsnames]{xcolor}
\usepackage{multirow}
\usepackage[colorlinks=true,citecolor=magenta,linkcolor=blue]{hyperref}
\usepackage[T1]{fontenc}
\usepackage{bbm}
\usepackage{thmtools,thm-restate}
\usepackage{verbatim}
\usepackage{mathtools}
\usepackage{titlesec}
\usepackage{amsmath}
\usepackage{subfigure}
\usepackage[normalem]{ulem}
\usepackage{circuitikz}

\usepackage{listings}
\usepackage{csquotes}

\newcommand{\dif}{\mathrm{d}}

\long\def\ca#1\cb{} 

\newcommand{\ket}[1]{|#1\rangle}               
\newcommand{\bra}[1]{\langle #1|}              

\newcommand{\dyad}[2]{\ket{#1}\!\bra{#2}}        

\renewcommand{\leq}{\leqslant}

\newcommand{\dt}{\mathrm{d}t}

\begin{document}

\title{Thermodynamic AI and the fluctuation frontier}
\author{Patrick J. Coles}
\author{Collin Szczepanski}
\author{Denis Melanson}
\author{Kaelan Donatella}
\author{Antonio J. Martinez}
\author{Faris Sbahi}
\affiliation{Normal Computing Corporation, New York, New York, USA}

\begin{abstract}

Many Artificial Intelligence (AI) algorithms are inspired by physics and employ stochastic fluctuations. We connect these physics-inspired AI algorithms by unifying them under a single mathematical framework that we call Thermodynamic AI. Seemingly disparate algorithmic classes can be described by this framework, for example, (1) Generative diffusion models, (2) Bayesian neural networks, (3) Monte Carlo sampling and (4) Simulated annealing.

Such Thermodynamic AI algorithms are currently run on digital hardware, ultimately limiting their scalability and overall potential. Stochastic fluctuations naturally occur in physical thermodynamic systems, and such fluctuations can be viewed as a computational resource. Hence, we propose a novel computing paradigm, where software and hardware become inseparable. Our algorithmic unification allows us to identify a single full-stack paradigm, involving Thermodynamic AI hardware, that could accelerate such algorithms.

We contrast Thermodynamic AI hardware with quantum computing where noise is a roadblock rather than a resource. Thermodynamic AI hardware can be viewed as a novel form of computing, since it uses a novel fundamental building block. We identify stochastic bits (s-bits) and stochastic modes (s-modes) as the respective building blocks for discrete and continuous Thermodynamic AI hardware. In addition to these stochastic units, Thermodynamic AI hardware employs a Maxwell's demon device that guides the system to produce non-trivial states. We provide a few simple physical architectures for building these devices and we develop a formalism for programming the hardware via gate sequences.

We hope to stimulate discussion around this new computing paradigm. Beyond acceleration, we believe it will impact the design of both hardware and algorithms, while also deepening our understanding of the connection between physics and intelligence.

\end{abstract}

\maketitle

\tableofcontents

\section{Introduction}

\subsection{The backstory}

In 2012, John Preskill wrote his landmark paper ``Quantum computing and the entanglement frontier''~\cite{preskill2012quantum}. In that paper, he gave a bird's-eye view of a new computing paradigm that is inherently based on the laws of physics pioneered by Peter Shor~\cite{shor1994algorithms}, Richard Feynman~\cite{feynman1982simulating} and many other visionaries. Preskill described entanglement both as something to be studied and as something to be utilized. In other words, it is both something that quantum computing will shed light on and something that is a necessary ingredient for building quantum computers. 

A decade later, quantum computing has turned into a tangible technology.  Superior performance relative to classical computers has been demonstrated on a specific problem~\cite{arute2019quantum, wu2021strong, zhong2021phase}~-- so-called quantum supremacy. However, despite tremendous progress, the field still remains in the Noisy-Intermediate Scale Quantum (NISQ) era~\cite{preskill2018quantum}, where ubiquitous commercial impact remains to be realized.

The timeline for quantum computing to reach large-scale commercial viability has been met with renewed skepticism and caution in many industrial and academic circles~\cite{waters_2023}. In part, this has resulted from discoveries such as the noise-induced barren plateau phenomenon~\cite{wang2020noise,franca2020limitations,wang2021can,bultrini2022battle,quek2022exponentially} that push applications like chemistry and optimization towards the so-called fault-tolerant era, and so further out in time. This is because noise can turn polynomial scaling algorithms into exponential scaling ones~\cite{wang2020noise,franca2020limitations}. Furthermore, many of the foundational principles and watershed applications of fault-tolerant quantum computing (qubits as a building block, unitary gates as a quantum program, quantum error correction and fault tolerance, exponential speedup for key applications like quantum simulation, etc.) have been worked out by pioneers of the field~\cite{nielsen2000quantum}. Nevertheless, intensive applied theoretical and engineering details remain to be sought.

Meanwhile alternative physics-based paradigms have been subject to increasing recent interest \cite{aadit2022massively,kaiser2022life,yamamoto2020coherent,vadlamani2020physics,mohseni2022ising,inagaki2016coherent,moy20221,chou2019analog,wang2019oim}. In many cases, these architectures have targeted combinatorial optimization problems (in particular, Quadratic Unconstrained Binary Optimization problems) with promising results~\cite{mohseni2022ising,inagaki2016coherent,moy20221,chou2019analog,wang2019oim}. Nonetheless, we will argue that alternative applications may offer a more natural mesh between the physics of the hardware and the algorithmic details of the application.

\subsection{A revolution needed}

Namely, there is a field that is making enormous commercial impact right now, a field that is highly technical and rich in theoretical problems, a field where noise not only might not be a major roadblock (unlike quantum computing) but actually might be utilized as part of the algorithm, a field where there is currently a mismatch between the software and the hardware, a field where many algorithms are inspired by physics (just like quantum computing), a field where there is an opportunity to develop a new hardware paradigm based on the laws of physics, and a field where the basic principles of that hardware paradigm have yet to be worked out. Specifically and broadly, this field is Artificial Intelligence~(AI). 

In the past few years, AI has exceeded the expectations of even many optimists. The release of text-to-image models in Stable Diffusion~\cite{rombach2021stable} and DALL-E~\cite{aditya2021dalle} made tremendous impact in the fields of automated art and automated design while capturing people's imagination. In addition, large language models that automate text generation have impacted many industries~\cite{radford2018improving}.

Much of this recent progress has happened through software and algorithmic advances, while largely keeping the standard hardware paradigm of digital computers with parallel processing. There may await another revolution in scaling up AI through fundamentally distinct, domain-specific hardware. This viewpoint has amassed increasing popularity, with initial strides where algorithm and hardware are considered inseparable~\cite{HintonNeurIPS2022}.

Undoubtedly, much of the explosion in AI progress over the last two decades has been driven by the scaling up of deep neural-network architectures. Meanwhile, the fundamental algorithmic components of such architectures had been in place for multiple decades prior to their recognition of unassailable promise. Indeed, a hardware ``fluke'' which presented equipment to accelerate matrix-vector multiplications (MVMs) enabled this step change \cite{hooker2021hardware}. Namely, this has been driven by parallelized digital hardware such as graphical processing units (GPUs) and field-programmable gate arrays (FPGAs)~\cite{kim2009highly}. More recently, analog hardware has been developed to reduce the power consumption in performing MVMs~\cite{du2018analog,wright2022deep}, and there is a rich history of analog hardware for neural networks~\cite{boser1991analog,sackinger1992application,verleysen1989analog}.

The next revolution in AI hardware may require identifying the physical basis of intelligence, possibly taking living systems as inspiration. For example, Jeremy England has connected many of the principles of life, such as self-organization, self-replication, and adaptation to fundamental ideas in thermodynamics such as the fluctuation-dissipation theorem~\cite{england2013statistical,england2015dissipative}. To the extent that intelligence is associated with life, then intelligence in turn could also be associated with thermodynamics. Independent of this view, some of the most successful recent approaches to Artificial Intelligence (AI) are inspired by physics, and often employ stochastic fluctuations.

\subsection{Symbiosis between fluctuations and intelligence}

Let us now elaborate on a potential symbiosis between thermodynamic fluctuations and intelligence.

We draw on the analogy to Preskill's proposed symbiosis between quantum computing and entanglement. Entanglement between multiple subsystems is, in general, difficult to simulate with classical computers and hence is a main factor in what gives quantum computers their power relative to classical computers. While entanglement is a key ingredient for quantum computers, studying it also is a key application for quantum computers, since for example topological quantum materials and strongly correlated-electronic systems such as superconductors typically exhibit many-body entanglement~\cite{amico2008entanglement} that quantum computers could shed light on.

By analogy, thermodynamic fluctuations could be a key ingredient for building (artificial) intelligence. Many AI primitives rely on simulating stochastic fluctuations~\cite{mansinghka2009natively}. Examples include generative diffusion models~\cite{song2020score}, time-series analysis with neural stochastic differential equations (SDEs)~\cite{kidger2020neural,li2020scalable} and Bayesian neural networks~\cite{goan2020bayesian,xu2022infinitely}. Generative modeling necessarily involves fluctuations since one does not simply want to replicate the data but rather randomly generate new samples. Instead of simulating such fluctuations with standard digital hardware, why not design a hardware that has thermodynamic fluctuations already built in as a fundamental building block.

Conversely large-scale thermodynamic AI machines could allow us to pursue the answers to deep questions about living systems or other adapting, self-evolving systems. There is a general principle in computing that one should simulate apples with apples~\cite{feynman1982simulating}. One should match the hardware design to the system that one would like to simulate, to maximize efficiency. There remain many interesting open questions about the origins of life, the origins of intelligent life, and the connections between life and the laws of physics, including testing England's theories related to these questions. Perhaps one could use thermodynamic AI machines as simulators for complex fluctuating systems. In this sense, thermodynamic AI machines will allow us to push forward into the fluctuation frontier.

\subsection{Goals of this article}

In writing this article, our first goal is to stimulate discussion around a new computing paradigm. We remark that Thermodynamic AI can be viewed as a sub-field of Thermodynamic Computing. The latter was discussed broadly in a workshop report~\cite{conte2019thermodynamic}, and was further explored by Hylton et al.~\cite{hylton2020thermodynamic,hylton2022thermodynamic}. Additional studies along these lines include thermodynamic neuromorphic systems~\cite{8123676,ganesh2020rebooting,8638594}, connecting machine learning to work production~\cite{Boyd_2022}, and other thermodynamic perspectives on learning~\cite{TherML,pmlr-v37-sohl-dickstein15,PhysRevLett.118.010601,doi:10.1146/annurev-conmatphys-031119-050745}. We use the term Thermodynamic AI to refer to a thermodynamic viewpoint on hardware for modern AI applications, and hence we hope to draw attention to this viewpoint.

Our second goal is to propose a possible mathematical and conceptual framework for Thermodynamic AI hardware. Indeed, in what follows we provide a framework that consists of:
\begin{enumerate}
\item Proposing stochasticity as a computational resource.

\item Identifying s-units (stochastic-units) as the basic building block of Thermodynamic AI hardware.

\item Discussing possible physical architectures for building s-units.

\item Developing a vector-space mathematical formalism to describe states, operators, and superoperators employed in Thermodynamic AI hardware.

\item Characterizing the gates that one can perform on single s-units and multiple s-units. 

\item Providing a mathematical framework for writing software programs to run on a Thermodynamic AI hardware in terms of gates sequences based on the aforementioned gates.

\item Connecting the thermodynamic concept of Maxwell's demon to AI algorithmic primitives., thus necessitating that Thermodynamic AI hardware include a physical realization of Maxwell's demon.

\item Presenting noise robustness and error correction concepts for Thermodynamic AI hardware.

\item Identifying key algorithms that could be sped up by Thermodynamic AI hardware, and unifying these algorithms under a single mathematical framework called Thermodynamic AI algorithms.
\end{enumerate}

\section{Stochasticity as a computing resource}\label{sc:stochasticity}

``Fluctuation'' is an intuitive term that is used by physicists oftentimes to describe the tendency to deviate from the average value. For example, vibrations of atoms in a crystal cause their spatial displacements to fluctuate about their equilibrium locations. We will essentially use the term ``fluctuation'' synonymously with the term ``stochasticity''.  Stochasticity has a precise mathematical description, and therefore it will allow us to precisely characterize the building blocks of thermodynamic AI systems. The key idea will be to view stochasticity as a resource that one can use to accomplish computational tasks. At first this may seem counter-intuitive, since stochasticity has inherent randomness to it.

However, randomness is useful as a resource for various tasks. Randomness is widely recognized as a resource by the cryptography community~\cite{hayes2001computing, gennaro2006randomness}. Randomness extraction protocols provide a crucial subroutine in cryptography, and random number generators are commercial technology. Randomness is also viewed as a resource in computing, including  quantum computing~\cite{liu2022noise}. This provides motivation for probabilistic computing based on p-bits~\cite{Camsari_2019,kaiser2022life,aadit2022massively,chowdhury2023full}. In particular, Monte Carlo algorithms, which represent one of the most widely used class of algorithms~\cite{kroese2014monte}, also employ randomness as a resource.

Naturally there is a close relationship between randomness and stochasticity, since these resources can be interconverted, albeit with some overhead. We remark that the idea that stochasticity is a computational resource has been promoted previously by Mansinghka~\cite{mansinghka2009natively}, in the context of digital stochastic circuits. We encourage the reader to see Ref.~\cite{mansinghka2009natively} for details on how stochasticity can accelerate various tasks, such as sampling problems. 

Let us illustrate how stochasticity, in particular, could be viewed as a resource. Consider the following examples:
\begin{itemize}
\item In generative modeling, it is common to add in noise to a target distribution because it makes the learning process easier. This is due to the manifold hypothesis~\cite{fefferman2016testing} (the data resides on a low dimensional manifold), which makes score matching poorly defined in the noise-free case. This has given rise to diffusion models~\cite{croitoru2022diffusion} whereby noise is digitally generated and added at discrete levels to the original data. Although effort has been made to make the noise levels more continuous, ultimately digital devices are inherently discrete. In contrast, a physical system with stochastic building blocks would naturally produce its own noise and would naturally evolve continuously. In this spirit of simulating apples with apples, the physical stochastic hardware would better match the underlying mathematics of diffusion models, and as a consequence one would expect better computational efficiency.

\item In annealing-based optimization algorithms such as simulated annealing, the (simulated) thermal fluctuations allow the system to explore different minima in the landscape before settling down in a high-quality minimum. If one was to use a physical system at finite temperature, then these thermal fluctuations would naturally occur, allowing one to explore many different minima in the landscape. Along these lines, researchers have indeed repurposed quantum annealers~\cite{nelson2022high} as physical thermal annealers to make use of these thermal fluctuations, although we remark that in principle thermal annealers do not need to be quantum devices.

\item Suppose that one wants to integrate a stochastic differential equation that describes the price of a financial asset. Digitally, one would discretize time and mimic stochasticity~\cite{oksendal2003stochastic} with pseudo-random discrete elements and then perform matrix-vector multiplications for a certain number of time steps to compute the integral. But if one had access to physical building blocks that were inherently stochastic, then one could appropriately set initial conditions and the parameters of their dynamics and then just check the state at a later time, without any computation at all and obtain the desired result.

\end{itemize}

\section{Unification of intelligent algorithms}\label{sc:unification}

\begin{figure}
    \centering
\includegraphics[width=0.8\textwidth]{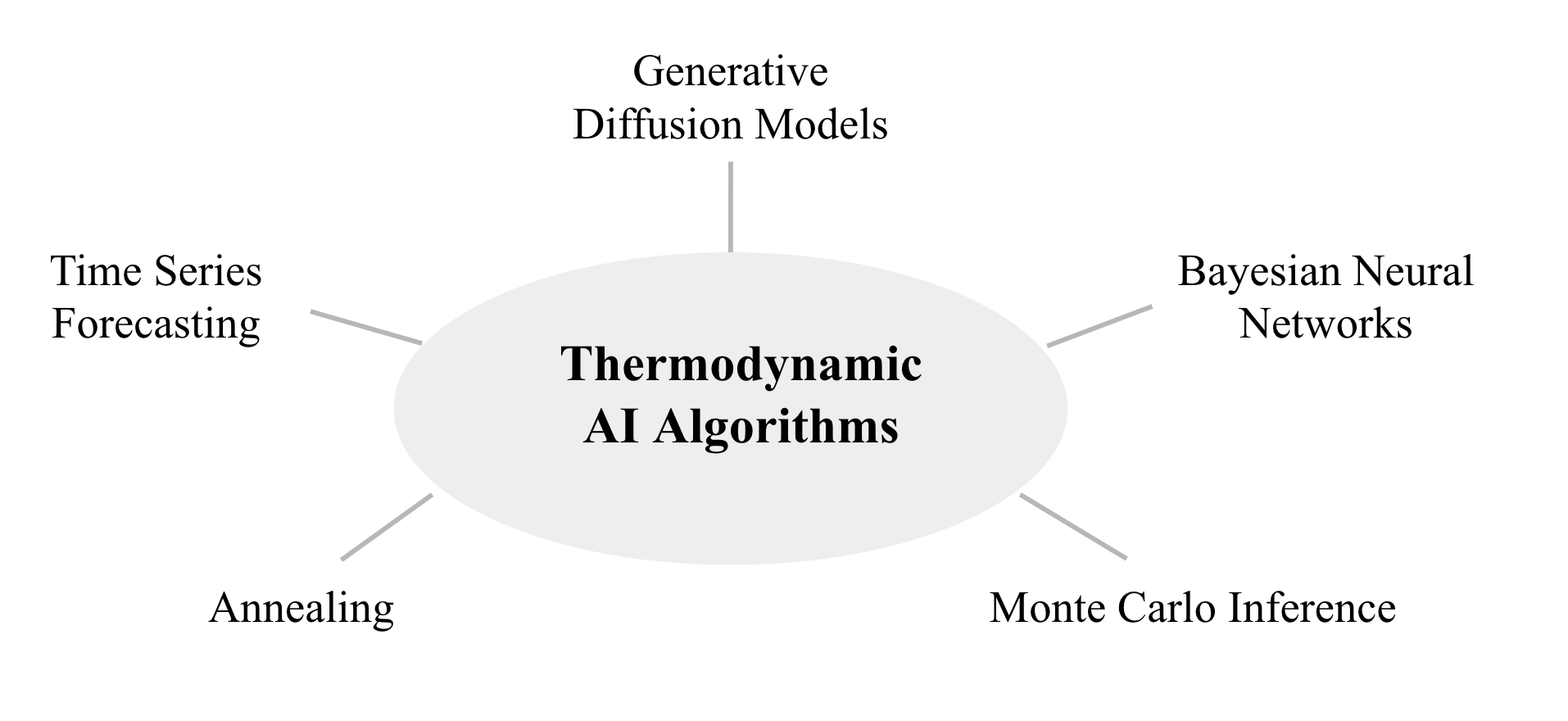}
    \caption{Illustration of various algorithms unified under the single mathematical framework of Thermodynamic AI algorithms.}
    \label{fig:Unified}
\end{figure}

In the field of physics, unifications are extremely powerful and highly sought after. For example, there used to be many different types of string theories, and then Ed Witten unified many different string theories under the umbrella of M-theory.

Our goal is to motivate a hardware paradigm that is relevant to multiple AI applications. A stepping stone towards this goal is to mathematically unify different AI algorithms under the same framework. Hence, a byproduct of our efforts is a conceptual and mathematical unification of AI algorithms that are generally considered different and unrelated. This is illustrated in Figure~\ref{fig:Unified}.

We alluded to some of these applications in the previous section, but let us reiterate them here:
\begin{enumerate}
\item Generative diffusion models (see Section~\ref{sc:diffusionmodels})
\item Bayesian neural networks (see Section~\ref{sc:TDL})
\item Monte Carlo inference (see Section~\ref{sc:montecarlo})
\item Annealing (see Section~\ref{sc:annealing})
\item Time series forecasting (see Section~\ref{sc:timeseries})
\end{enumerate}
Each application is discussed in more detail later in the article, with the relevant sections indicated.

Through careful consideration, we manage to formulate a mathematical framework that encompasses all of the aforementioned algorithms as special cases. We say that these algorithms belong to a class called \textit{Thermodynamic AI algorithms}. 

At a conceptual level, we can define Thermodynamic AI algorithms as algorithms consisting of at least two subroutines:
\begin{enumerate}
    \item  A subroutine in which a stochastic differential equation (SDE) is evolved over time.
    \item A subroutine in which a Maxwell's demon (see Sec.~\ref{sc:entropy_MD} for elaboration) observes the state variable in the SDE and applies a drift term in response.
\end{enumerate}

At the mathematical level, we propose that Thermodynamic AI algorithms are ones that simulate or implement the following set of equations (or some subset of them):
\begin{align}
d\mathbf{p} &= [\mathbf{f} - BM^{-1}\mathbf{p}]\hspace{1pt}dt + D \hspace{1pt}d\mathbf{w}\\
d\mathbf{x} &= M^{-1}\mathbf{p}\hspace{1pt}dt \\
\mathbf{f}&= - \nabla_{\mathbf{x}} U_{\theta}
\end{align}
One can see that these correspond to Newton’s laws of motion, with the addition of diffusion and friction. In these equations, $\mathbf{p}$, $\mathbf{x}$, and $\mathbf{f}$ respectively are the momentum, position, and force. The matrices $M$, $D$, and $B$ are hyperparameters, with $M$ being the mass matrix and $D$ being the diffusion matrix. The $d\mathbf{w}$ term is a Wiener process. Finally, $U_{\theta}$ is a (trainable) potential energy function. Typically, much of the application-specific information, regarding the task to be solved, is encoded in the potential energy function $U_{\theta}$. Note that for readability we omit the dependencies of the variables on time $t$ and space $\mathbf{x}$ in the above equations.

As we said, this unification is crucial to developing a hardware paradigm that is broadly applicable to many AI algorithms. However, the mathematical unification can itself be useful, even outside of the development of novel hardware. All of the aforementioned algorithms are currently implemented in standard, digital hardware, but typically with different software programs. In principle, one can use the mathematical framework presented in this article to develop a unified algorithmic framework for running these applications on standard, digital hardware. For example, see Refs.~\cite{goodman2012church,cusumano2019gen} for early work on a unified software approach for probabilistic algorithms.

In this article we aim to keep the focus on our novel hardware paradigm. Hence, we simply leave this algorithmic unification as a remark that we do not discuss further in this article. On the other hand, we do discuss the idea of unified software for programming Thermodynamic AI hardware later in the article, and our mathematical unification is important for this purpose.

\section{Fundamental building blocks}

\begin{center}
\begin{tabular}{|c|c|c|c|c| }
 \hline
  & classical & quantum & probabilistic & thermodynamic \\ 
 \hline
discrete & bit &qubit & p-bit & s-bit \\ 
continuous & mode & qumode & p-mode & s-mode \\ 
 \hline
\end{tabular}
\end{center}

Let us now discuss the fundamental building blocks of thermodynamic AI hardware. As the name ``thermodynamic'' suggests, a thermodynamic system is inherently dynamic in nature. Therefore, the fundamental building blocks should also be dynamic. This is contrast to classical bits or qubits, where the state of the system ideally remains fixed unless it is actively changed by gates. The thermodynamic building block should passively and naturally evolve over time, even without the application of gates.

But what dynamical process should it follow? A reasonable proposal is a stochastic Markov process. Naturally this should be continuous in time, since no time point is more special than any other time point. Hence, the discrete building block, which we call an s-bit, would follow a continuous-time Markov chain (CTMC). Here the ``s'' in s-bit stands for stochastic.

For the continuous building block, which we call an s-mode, the natural analog would be a Brownian motion (also known as a Wiener process). One can impose that this process is a Martingale (which is typically assumed for Brownian motion), which means that it has no bias.  We use s-unit as a generic term to encompass both s-bits and s-modes. We note that stochastic building blocks were also considered by Mansinghka et al.~\cite{mansinghka2009natively,mansinghka2013combinational} using digital logic, which is different from our (analog) approach. 

For comparison, one can consider the fundamental building blocks of a probabilistic system, which are p-bits (p-modes) for the discrete case (continuous case). The p-bit can be thought of as a random number generator, which either generates 0 or 1 at random. The analog of this in the continuous case, which we call a p-mode, could be a random number generator that generates a real number according to a Gaussian distribution with zero mean and some variance. 

Although p-bits (p-modes) are clearly different than s-bits (s-modes), there is some connection between them. Namely, one could think of the latter as the time integral of the former. For example, one could view an s-mode as the Ito integral of a p-mode. In this sense, one could experimentally construct s-bits (s-modes) if one has access to both p-bits (p-modes) as well as a time-integrator apparatus.

\section{Physical realizations of s-units}\label{sc:physical_realization}

\subsection{Physical realizations of s-modes}

\subsubsection{Individual s-modes}\label{sc:phys_smode}

An s-mode represents a continuous stochastic variable whose dynamics are governed by drift, diffusion, or other physical processes (akin to a Brownian particle). At the heart of any physical implementation of such a variable will be a source of stochasticity. A natural starting point for implementing thermodynamic AI hardware is analog electrical circuits, as these circuits have inherent fluctuations that could be harnessed for computation.  

The most ubiquitous source of noise in electrical circuits is thermal noise \cite{horowitz1989art}. Thermal noise, also called Johnson-Nyquist noise, comes from the random thermal agitation of the charge carriers in a conductor, resulting in fluctuations in voltage or current inside the conductor. Thermal noise is Gaussian and has a flat frequency spectrum (white noise) with fluctuations in the voltage of standard deviation
\begin{align}
    v_{\mathrm{tn}} = \sqrt{4 k_B T R \Delta f},
\end{align}
where $R$ is the resistance of the conductor, $k_B$ is the Boltzmann constant, $T$ is the absolute temperature and $\Delta f$ is the frequency bandwidth. The amplitude of the voltage fluctuations can be controlled by changing the temperature or the resistance. In practice, a thermal noise source can be implemented using a large resistor in series with a voltage amplifier.

Another type of electrical noise is shot noise \cite{horowitz1989art}. Shot noise arises from the discrete nature of charge carriers and from the fact that the probability of a charge carrier crossing a point in a conductor at any time is random. This effect is particularly important in semiconductor junctions where the charge carriers should overcome a potential barrier to conduct a current. The probability of a charge carrier passing over the potential barrier is an independent random event. This induces fluctuations in the current through the junction. Shot noise is Gaussian and has a flat frequency spectrum (white noise) with fluctuations in the current of standard deviation 
\begin{align}
    I_{\mathrm{tn}} = \sqrt{2 q |I| \Delta f},
\end{align}
where $I$ is the current through the junction, $q$ is the electron charge and $\Delta f$ is the frequency bandwidth. The amplitude of the current fluctuations can be controlled by changing the magnitude of the DC current passing through the junction. In practice, a source of shot noise would be implemented using a $pn$ diode (for example, a Zener diode in reverse bias configuration) in series with a controllable current source~\cite{lecoy1972noise}.

In general, any physical implementation of s-modes should have the amplitude of its stochasticity be independently controllable with respect to the other system parameters. We have seen that electrical thermal and shot noise, for example, both have tuning knobs to control the amplitude of the noise to some extent. 

In addition, one must ensure that the amplitude of the fluctuations be compatible, i.e. measurable, with the rest of the system. Thermal and shot noise sources typically have voltage fluctuations of the order of a few $\mu V$ or less. For these fluctuations to be measurable by on-chip analog-to-digital converters measuring voltages on the order of hundreds of $m V$, amplification will be necessary. This amplification can be done using single- or multi-stage voltage amplifiers. Variable-gain amplifiers can also let one independently control the amplitude of the fluctuations.  

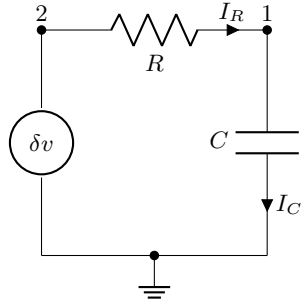
\begin{figure}
    \centering
    \begin{circuitikz}
        \draw (0,1.5) node[rmetershape](dv){} node[]{$\delta v$};
        \draw (0,0) -- (dv) to[short, -*] (0,3) node[above]{$2$};
        \draw (0,3) to[R, l_=$R$, i=$I_{R}$, -*] (3,3) node[above]{$1$};
        \draw (3,3) to[C, l_=$C$, i=$I_{C}$] (3,0);
        \draw (0,0) to[short, -*] (1.5,0) node[ground]{} -- (3,0);
    \end{circuitikz}
    \caption{Circuit diagram of a possible physical realization of an s-mode, comprising of a noisy resistor and a capacitor.}
    \label{fig:smode_rc}
\end{figure}

The s-mode can be represented through the dynamics of any degree of freedom of an electrical circuit. If we chose the voltage on a particular node in the circuit as our degree of freedom of choice, a simple stochastic voltage noise source plays the role of the s-mode. This can be realized by using a noisy resistor at non-zero temperature. The circuit schematic in Fig. \ref{fig:smode_rc} shows the typical equivalent noise model for a noisy resistor composed of a stochastic voltage noise source, $\delta v (t)$, in series with an ideal (non-noisy) resistor of resistance $R$. The inherent terminal capacitance, $C$, of the resistor is also added to the equivalent resistor model \cite{vanzon2004fluctuations, vasziova2010thermal}. We chose to use the voltage on node 1 (labeled simply as $v (t)$ here) as our s-mode. The dynamics of the s-mode in this case, obeys the following SDE model:
\begin{align} \label{eq:smode_rc}
    -\frac{\mathrm{d}v (t)}{\mathrm{d}t} = \frac{v (t) + \delta v (t)}{RC}.
\end{align}
The voltage fluctuations in Eq. (\ref{eq:smode_rc}) have the usual Gaussian white noise properties $\langle \delta v (t) \rangle = 0$ and $\langle \delta v (t) \delta v (t') \rangle = 2 k_B T R \delta (t - t')$ where, $\langle~\rangle$ represents statistical averaging and where $\delta (t - t')$ represents the Dirac delta function. We notice the form of the SDE comprises a drift term proportional to $v (t)$ and a diffusion or stochastic term proportional to $\delta v (t)$.  

In order for the s-modes to be more useful for computation, their inherent stochastic dynamics must be constrained to the properties of an algorithm. One can, for example add a drift term to the dynamics of the s-mode by adding a capacitor in series with the noisy resistor. In general, one can add other electrical components, such as inductors or non-linear elements, to further constrain the evolution of the s-mode.

\subsubsection{Coupling s-modes}
When building systems of many s-modes, one will most likely wish to introduce some form of coupling between them to express correlations and geometric constraints. Again, the medium of analog electrical circuits presents a natural option for the coupling of s-modes. 

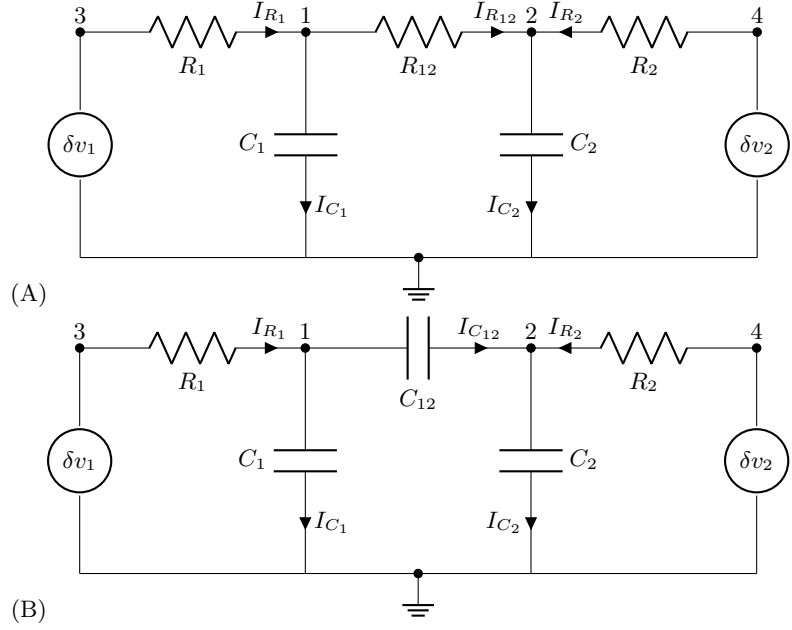
\begin{figure}[t]
    \centering
    (A)\begin{circuitikz}
        \draw (0,1.5) node[rmetershape](a1){} node[]{$\delta v_1$};
        \draw (0,0) -- (a1) to[short, -*] (0,3) node[above]{$3$};
        \draw (0,3) to[R, l_=$R_{1}$, i=$I_{R_1}$, -*] (3,3) node[above]{$1$};
        \draw (3,3) to[C, l_=$C_{1}$, i=$I_{C_1}$] (3,0);
        \draw (3,3) to[R, l_=$R_{12}$, i=$I_{R_{12}}$, -*] (6,3) node[above]{$2$};
        \draw (6,0) to[C, l_=$C_{2}$, i<=$I_{C_2}$] (6,3);
        \draw (6,3) to[R, l_=$R_{2}$, i<=$I_{R_2}$] (9,3);
        \draw (9,1.5) node[rmetershape](a2){} node[]{$\delta v_2$};
        \draw (9,0) -- (a2) to[short, -*] (9,3) node[above]{$4$};
        \draw (0,0) to[short, -*] (4.5,0) node[ground]{} -- (9,0);
    \end{circuitikz}
    (B)\begin{circuitikz}
        \draw (0,1.5) node[rmetershape](a1){} node[]{$\delta v_1$};
        \draw (0,0) -- (a1) to[short, -*] (0,3) node[above]{$3$};
        \draw (0,3) to[R, l_=$R_{1}$, i=$I_{R_1}$, -*] (3,3) node[above]{$1$};
        \draw (3,3) to[C, l_=$C_{1}$, i=$I_{C_1}$] (3,0);
        \draw (3,3) to[C, l_=$C_{12}$, i=$I_{C_{12}}$, -*] (6,3) node[above]{$2$};
        \draw (6,0) to[C, l_=$C_{2}$, i<=$I_{C_2}$] (6,3);
        \draw (6,3) to[R, l_=$R_{2}$, i<=$I_{R_2}$] (9,3);
        \draw (9,1.5) node[rmetershape](a2){} node[]{$\delta v_2$};
        \draw (9,0) -- (a2) to[short, -*] (9,3) node[above]{$4$};
        \draw (0,0) to[short, -*] (4.5,0) node[ground]{} -- (9,0);
    \end{circuitikz}
    \caption{Circuit diagram of possible physical realizations of coupling between s-modes, using (A) a coupling resistor and (B) a coupling capacitor.}
    \label{fig:phys_RandCcoupled_smodes}
\end{figure}

As a first example, two circuits of the type described in section \ref{sc:phys_smode} could be coupled through a resistor, as pictured in the upper panel of Fig. \ref{fig:phys_RandCcoupled_smodes}. The coupled s-modes, represented by the voltage on nodes 1 and 2, are then coupled through their drift terms as (we omit the time dependencies for readability)
\begin{align}\label{eq:Rcoupled_smodes}
    -\frac{\dif v_1}{\dif t} &= \frac{v_1 + \delta v_1}{R_1 C_1} - \frac{v_2 - v_1}{R_{12} C_1}, \\
    -\frac{\dif v_2}{\dif t} &= \frac{v_2 + \delta v_2}{R_2 C_2} - \frac{v_1 - v_2}{R_{12} C_2}.
\end{align}
To simplify the notation, this system of coupled equations can be written in matrix form as follows
\begin{align}\label{eq:Rcoupled_smodes_matrix}
    -\mathbf{\dot{v}} = \mathbf{C}^{-1} \left(\mathbf{J}\mathbf{v} + \mathbf{R}^{-1} \mathbf{\delta v}\right),
\end{align}
where $\mathbf{\dot{v}} \equiv \frac{\dif}{\dif t} \mathbf{v}$ and
\begin{align*}
    \mathbf{v} \equiv
    \begin{bmatrix}
        v_{1}\\
        v_{2}
    \end{bmatrix},~
    \mathbf{C}\equiv 
    \begin{bmatrix}
        C_1 & 0\\
        0 & C_2 
    \end{bmatrix},~
    \mathbf{J} \equiv 
    \begin{bmatrix}
        \frac{1}{R_1} + \frac{1}{R_{12}} & -\frac{1}{R_{12}}\\
        -\frac{1}{R_{12}} & \frac{1}{R_2} + \frac{1}{R_{12}} 
    \end{bmatrix},~
    \mathbf{R} \equiv 
    \begin{bmatrix}
        R_1 & 0\\
        0 & R_2 
    \end{bmatrix},~
    \mathbf{\delta v} \equiv 
    \begin{bmatrix}
        \delta v_1\\
        \delta v_2
    \end{bmatrix}.
\end{align*}
Where we have introduced the self-resistance matrix $\mathbf{R}$, the capacitance matrix $\mathbf{C}$, and the conductance matrix $\mathbf{J}$.

A second method of coupling two s-modes together is by using a capacitor as a the coupling element, as pictured in the lower panel of Fig. \ref{fig:phys_RandCcoupled_smodes}. This coupling scheme has been studied in the context of heat engines and entropy production \cite{Ciliberto2013heat, chiang2017entropy}, where the voltage in both cells was found to be correlated for large capacitive coupling. In this configuration, the two coupled s-modes, represented by the voltage on nodes 1 and 2, have drift and diffusion coupling as (we omit the time dependencies for readability)
\begin{align}\label{eq:Ccoupled_smodes}
    -\frac{\dif v_1}{\dif t} &= \frac{v_1 + \delta v_1}{R_1(C_1 + C_{12})} + \frac{C_{12} (v_2 + \delta v_2)}{R_2(C_1 + C_{12}) (C_2 + C_{12})} \\
    -\frac{\dif v_2}{\dif t} &= \frac{v_2 + \delta v_2}{R_2 (C_2 + C_{12})} + \frac{C_{12} (v_1 + \delta v_1)}{R_1 (C_1 + C_{12}) (C_2 + C_{12})}.
\end{align}
To simplify the notation, using the same notation as in Eq.~\eqref{eq:Rcoupled_smodes_matrix}, this system of coupled equations can be written in matrix form as follows
\begin{align}\label{eq:Ccoupled_smodes_matrix}
     -\mathbf{\dot{v}} = \mathbf{C}^{-1} \mathbf{R}^{-1} \left(\mathbf{v} + \mathbf{\delta v}\right),
\end{align}
where the capacitance matrix is
\begin{align*}
    \mathbf{C} \equiv 
    \begin{bmatrix}
        C_1 + C_{12}& -C_{12}\\
        -C_{12} & C_2 + C_{12} 
    \end{bmatrix}.
\end{align*}

\subsection{Physical realizations of s-bits}

\subsubsection{Individual s-bits}

Recall that an s-bit evolves as a sample trajectory $x(t)$ of a continuous-time Markov chain (CTMC) over $\{0,1\}$. A program begins by initializing an s-bit state $x_0 \in \{0,1\}$. The state of the s-bit will be stored in digital memory, but will be flipped to the opposite state at random instances triggered by a physical noisy system. In particular, we can leverage DC analog components that release shot noise, and let the timing of these events dictate when s-bit flips occur. Shot noise emitted from a physical component can be modeled by a Poisson process, with some experimentally-determined rate constant $\lambda$ intrinsic to the analog component producing the noise.
If we flip the value of the s-bit in memory whenever shot noise is emitted, the time between s-bit flips will be approximately distributed according to an exponential distribution with the rate constant $\lambda$ intrinsic to the physical noise source. If we know the initial state $x_0$ of the s-bit, knowing the time of all shot events up to time $t>0$ is sufficient to tell the state of the s-bit $x(t)$ at all times up to t.

The s-bit, randomly flipping states at times sampled from exponential distributions by the physical shot noise source, undergoes a CTMC over the pair of discrete $\{0,1\}$ states. The value $x(t)$ of the s-bit state throughout a program is then a sample trajectory of this Markov chain. Using a single shot noise source to indicate bit flips, shot events will flip the average s-bit from $0 \to 1$ as quickly as it flips it from $1 \to 0$. These events happen at a rate $\lambda$ that is intrinsic to the physical noise source, and the s-bit follows a trajectory of a CTMC with symmetric rate matrix $Q$: 
\begin{equation} \label{sbit_matrix}
   Q = \begin{pmatrix}
1-\lambda & \lambda \\
\lambda & 1-\lambda \\
\end{pmatrix}
\end{equation}
As a Markov chain is a dynamic process with a time dimension, we also suppose the parameters of the physical noise source can depend on time, so that the rate $\lambda$ in \eqref{sbit_matrix} can be made to depend on time according to some user-specified schedule $\lambda(t)$; the rate at which flip events occur is dependent on how long the program has been running for. With these modifications, one can program s-bit dynamics with increased complexity, but the rates of hopping from $0 \to 1$ and vice versa always remain equal, as the physical noise source does not "update" its behavior based on the current state of the s-bit. We allow for CTMC with non-symmetric rate matrices by establishing a feedback loop between the noise source and the current state of the s-bit.

In order to simulate a Markov chain with a non-symmetric rate matrix, the rate schedule realized by the physical Poisson process must change when the s-bit state changes. If the s-bit jumps to the value 0 at time $\tau$, we program the noise source to produce shot events with a rate schedule $\lambda_0(t)$, from time $\tau$, until the first shot event happens at some time $T > \tau$.

The shot noise indicates a flip of the s-bit state from 0 to 1 at time $T$; immediately, we change the rate schedule of the physical Poisson process and realize a new schedule $\lambda_{1}(t)$ from time $T$ until the next shot event happens at a later time $\mathcal{T} > T$ with a new applied current $I_{1}(t)$. The region of the rate schedule we are interested in changes as the program advances in time, and this update scheme requires a feedback loop between the state of the s-bit $x(t)$ and the parameters of the physical noise device. If we repeat this procedure, the state of the s-bit simulates a sample trajectory of the Markov chain with time-dependent rate matrix $Q(t)$: 
\begin{equation} \label{sbit_matrix2}
  Q(t) =   \begin{pmatrix}
1-\lambda_{0}(t) & \lambda_{0}(t) \\
\lambda_{1}(t) & 1-\lambda_{1}(t) \\
\end{pmatrix}
\end{equation}

\subsubsection{Coupling s-bits}

S-bits may be coupled together, and dynamics of the joint system can simulate Markov processes over any data space that can be represented with a finite encoding scheme. A system of $N$ coupled s-bits will hold a state $x(t) \in \{0,1\}^N$ that changes in time during a program; as in the case of a single s-bit, the system of coupled s-bits will evolve as a sample trajectory from a user-specified CTMC. 

A user specifies a Markov chain by choosing a $2^N \times 2^N$ dimensional rate matrix $Q(t)$, which collects the time-dependent rates $q_{xy}(t)$ of transitions between all pairs of states $x,y \in \{0,1\}$. Note that the rate of transitions $x \to y$ is labeled $q_{xy}(t)$, while the rate of transitions $y \to x$ is labeled $q_{yx}(t)$; these rate schedules need not be equal. For an arbitrary state $x$, the possible transitions are the transitions to states $y$ with non-zero rate schedules $q_{xy}(t)$; let $n_x$ denote the number of possible transitions out of $x$. 

In the case of a single s-bit, only one possible transition exists out of either state the s-bit can hold. In the coupled case, the number of possible transitions from any state $x \in \{0,1\}^N$ will be higher and a noise source for each of these possible transitions will be required. As with single s-bits, a feedback loop between the state of the coupled s-bits $x(t)$ and the noise sources must exist to re-program the parameters of the noise sources every time the s-bit system transitions to a new state, for a new set of possible transitions out of that state exist. 

Suppose that the s-bit system $x(t)$ jumps to state $x \in \{0,1\}^N$ at time $\tau$. For each state $y \in \{0,1\}^N$ that the s-bit in state $x$ is allowed to transition to, we program a physical noise source to produce shot noise with rate schedule $q_{xy}(t)$ beginning at time $\tau$. To accomplish this, we apply a current schedule $I_{xy}(t)$, beginning from $I_{xy}(\tau)$, to each diode representing the $x \to y$ transitions, such that each noise source produces noise a rate schedule $q_{xy}(t)$ from the correct starting time. 

If there are $n_x$ such possible transitions, the s-bit system $x(t)$ in state $x$ will jump to a new state according to the time $T >0$ and index $y'$ of the noise source that emitted a shot event first out of $n_x$ "competing" noise sources. At this time $T$, we update the value of the s-bit system $x(T) = y'$, and reset the physical noise sources; they will now be re-used to sample transitions out of the new state $y'$, beginning at a new time T seconds into the rate schedules. 

With this update scheme and a rule for selecting initial s-bit states $x_0$, the value $x(t)$ of the s-bit system during a program is a sample trajectory of a continuous time Markov chain over $\{0,1\}^N$ according to the $2^N \times 2^N$ time-dependent rate matrix $Q(t)$.

\subsection{Problem geometry, inductive bias, and connectivity}

\begin{figure}
    \centering
(A)\includegraphics[width=0.46\textwidth]{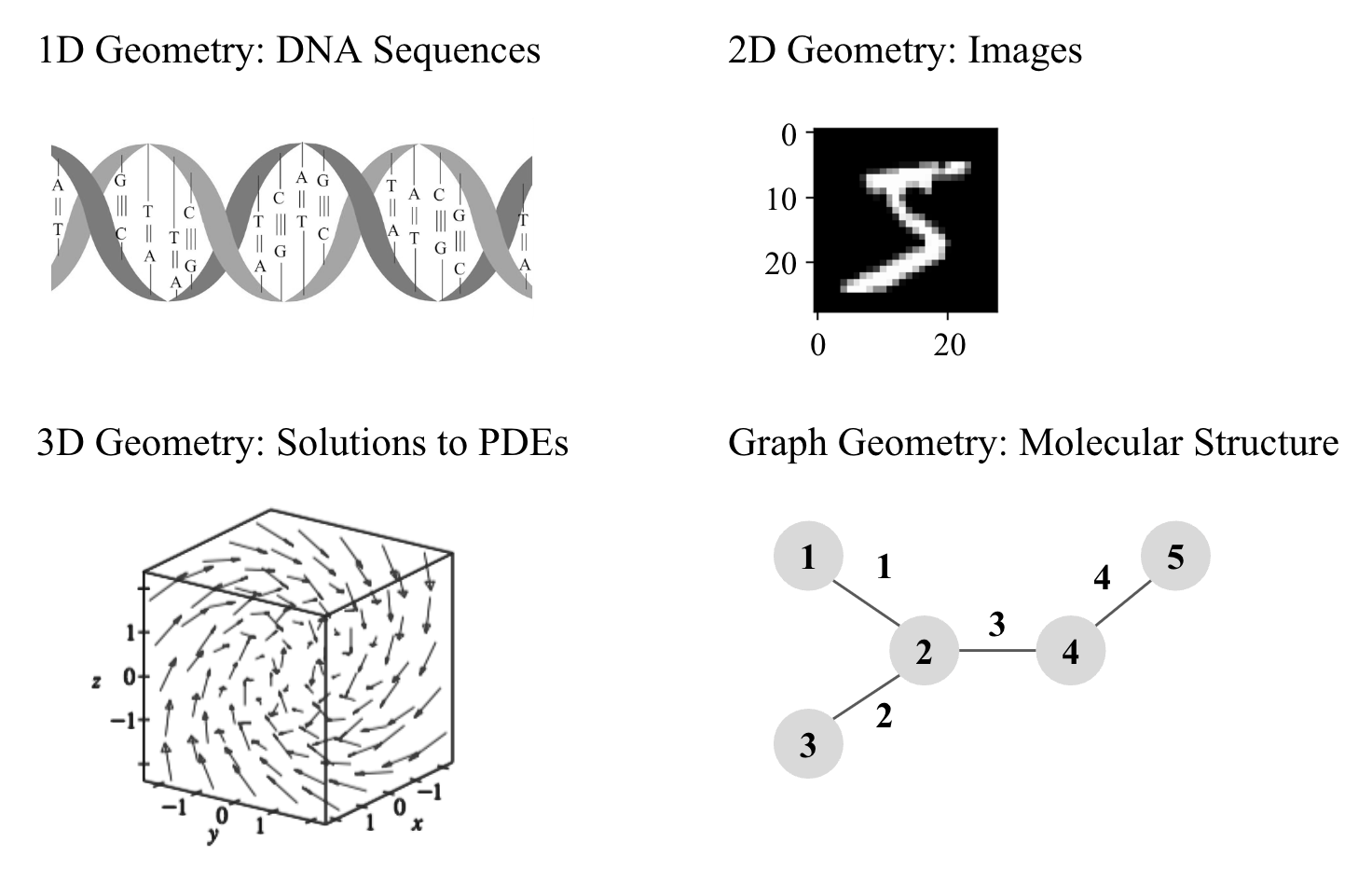}
(B)\includegraphics[width=0.46\textwidth]{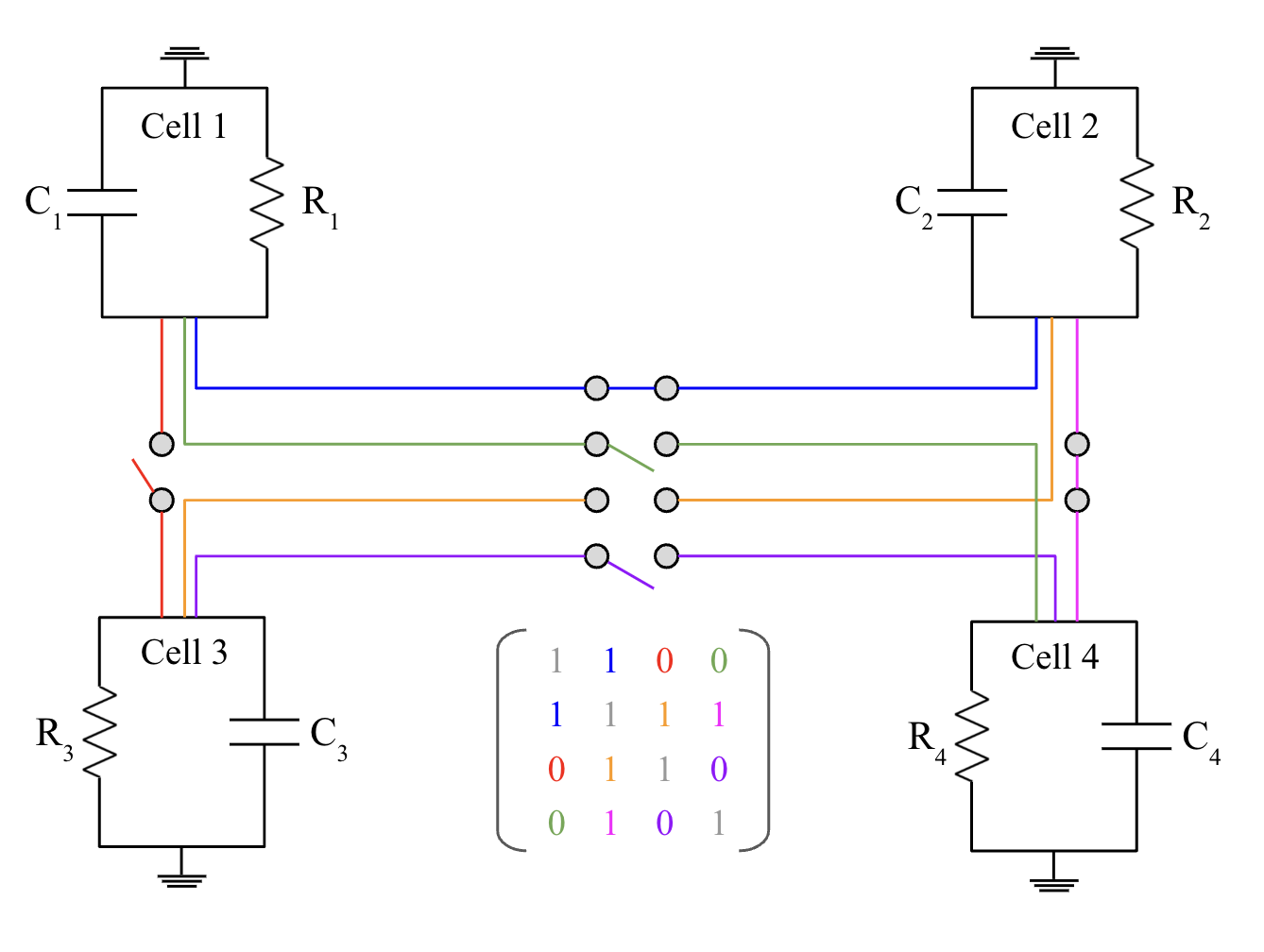}
    \caption{Accounting for problem geometry in designing and programming Thermodynamic AI Hardware. (A) Problems can have different geometries, such as 1D, 2D, 3D, and graph geometries. Incorporating geometric information often improves the training and generalization performance of machine learning models, providing an inductive bias. One can take the same approach of adding an inductive bias to Thermodynamic AI systems. (B) We show that switches can be added to the connections between s-units. The geometry of the problem is mathematically summarized in an adjacency matrix, and this adjacency matrix can determine the connectivity of the s-units. One can choose the connectivity either at the hardware level (hard-wired connectivity) or at the software level (software controlled connectivity).}
    \label{fig:Connectivity}
\end{figure}

Figure~\ref{fig:Connectivity} illustrates how one can account for problem geometry in designing and programming Thermodynamic AI Hardware.

In the left panel, we provide examples of problems that have different geometries. DNA sequences and time-series data are 1D, images are 2D, crystalline materials and solutions are partial differential equations are often 3D, and molecules can be represented as graphs.

Accounting for geometric information can improve the training and generalization performance of machine learning models, by providing an inductive bias. Hence one can take the same approach of adding an inductive bias to Thermodynamic AI systems. Inductive bias allows one to limit the search space during the training process and oftentimes speeds up training.

Mathematically, geometry is often translated into the language of group theory. Typically one aims for invariance or equivariance of the machine learning model under group element action~\cite{bronstein2021geometric,larocca2022group,nguyen2022theory}. Indeed, constructing Thermodynamic AI hardware to be equivariant for specific symmetry groups is possible, although we do not delve into the details. 

For simplicity, we illustrate that geometric information can be summarized in terms of an adjacency matrix, such as the matrix in the right panel of Figure~\ref{fig:Connectivity}. This abstract adjacency matrix can be translated into a concrete connectivity for the Thermodynamic AI hardware. Namely, switches can be added to bridges that connect the s-units. The state of these switches (whether they are open or closed) can be determined by the matrix elements of the  adjacency matrix. 

We remark that one can choose the connectivity either at the hardware level (hard-wired connectivity) or at the software level (software-controlled connectivity). Hard-wiring the connectivity would make the hardware relevant to applications with a specific geometry, e.g., one could hard-wire a 2D connectivity for image applications. Software-controlled connectivity provides the flexibility to use the same hardware for applications with different geometries.

\section{Thermodynamic AI Hardware as 
a hybrid digital-analog system}

We remark that, in practice, Thermodynamic AI hardware will typically be a hybrid system of digital and analog components. 

The s-bit or s-mode device represents the key analog component of the overall system. As discussed in Sec.~\ref{sc:physical_realization}, this device will be constructed from physical analog sub-units, whose time evolution are governed by physical processes.

As we will see later in Sec.~\ref{sc:entropy_MD}, the s-unit device is not the only component of Thermodynamic AI hardware. In that section, we will discuss the necessity of additional hardware that is coupled to the s-unit device. This additional hardware will allow the overall system to produce complex probability distributions and complex entropy dynamics. The additional hardware could be digital (as shown later in Fig.~\ref{fig:ScoreNetworkMD}), or it could be analog or hybrid digital-analog as discussed in Sec.~\ref{sc:physical_MD}. We will aim to highlight the benefits and weaknesses of certain components being digital or analog. 

Nevertheless, we propose for thermodynamic AI systems to in fact be hybrid digital-analog systems, where the fraction of digital hardware may be changed depending on the task at hand.

\section{States, operators, and superoperators}

In order to describe the dynamics of s-units, one first needs to describe the state of the system. Ideally one would like to leverage the mathematical formalism of vector spaces, and hence we propose the state of an s-unit lives in a vector space.

\subsection{Vector space formalism for s-modes}\label{sc:smode_formalism}

\subsubsection{State space}

Let us first focus on the case of s-modes. In this case, we imagine a system with $N$ s-modes, where each s-mode has a continuous state variable describing its state at time $t$. Let $v_j(t)\in \mathbb{R}$ be a real number, and let it be the continuous state variable associated with the $j$-th s-mode, with $j = 1, ..., N$. Then, the state vector can be written as a column vector:
\begin{equation}\label{eqn_v_1}
    \mathbf{v}(t) = \{v_1(t), v_2(t), ..., v_N(t)\}^T
\end{equation}

The vector $\mathbf{v}(t)$ belongs to an $N$-dimensional vector space, which we call the state space. The state space is a vector space over the field of real numbers. Formally speaking, we can denote the state space as $\mathcal{S}$, and we can write:
\begin{equation}\label{eqn_v_2}
    \mathbf{v}(t) \in \mathcal{S}\,.
\end{equation}

\subsubsection{Operator space}

The next question is how $\mathbf{v}(t)$ evolves over time. A simple model would be a stochastic differential equation with both a drift term and a diffusion term, such as the following:
\begin{equation}\label{eqn_generalevolution_smodes}
    d\mathbf{v}(t) = (A(t) \mathbf{v}(t)+\mathbf{b}(t))dt + C(t) d\mathbf{w}
\end{equation}
Here $A(t)$ and $C(t)$ are $N\times N$ matrices, and $\mathbf{b}(t)$ is an $N$-dimensional vector, all with entries that are real numbers. We respectively call $A(t)$, $\mathbf{b}(t)$, and $C(t)$ the drift matrix, the drift vector, and the diffusion matrix. The $d\mathbf{w}$ term represents an $N$-dimensional Brownian motion (also known as Wiener process).

$A(t)$ and $C(t)$ belong to a vector space of their own, which we call the operator space. The operator space is $N^2$-dimensional, and again is defined over the field of real numbers. The operator space can be viewed as the space of linear operators $\mathcal{L}(\mathcal{S})$ acting on the state space $\mathcal{S}$. Hence we can write:
\begin{equation}
    A(t), C(t) \in \mathcal{L}(\mathcal{S}),\quad\text{and}\quad \mathbf{b}(t) \in \mathcal{S}\,.
\end{equation}

In order to define the bare-bones hardware (i.e., the set of $N$ s-modes without any gates applied), one can take $A(t)$, $\mathbf{b}(t)$, and  $C(t)$ to be fixed, pre-determined functions. We can write $A(t) = A_0(t)$, $\mathbf{b}(t) = \mathbf{b}_0(t)$, and $C(t) = C_0(t)$ for these pre-determined functions. In other words  $A_0(t)$, $\mathbf{b}_0(t)$, and $C_0(t)$ represent the bare-bones drift matrix, drift vector, and diffusion matrix associated with the set of s-modes as they are physically constructed (i.e., without any gates applied).

For notational simplicity, let us assume that these bare-bones objects are constant in time, and respectively denote them as $A_0$, $\mathbf{b}_0$, and $C_0$. Under this assumption, we can write the SDE for the bare-bones hardware (i.e., with no gates applied) as
\begin{equation}\label{eqn_barebones_smode}
    d\mathbf{v}(t) = (A_0 \mathbf{v}(t)+\mathbf{b}_0)dt + C_0 d\mathbf{w}\,.
\end{equation}
We remark that, in practice, one might choose $A_0$, $\mathbf{b}_0$, and $C_0$ to be particularly simple. For example, setting $\mathbf{b}_0 = \mathbf{0}$ and choosing the matrices to proportional to the identity, $A_0 = a_0 I $ and $C_0 = c_0 I $, would be a reasonable choice in order to minimize the physical overhead of constructing the hardware.

\subsubsection{Superoperator space}\label{sc:superoperator}

Let us now describe how to go from the bare bones hardware in Eq.~\eqref{eqn_barebones_smode} to the general evolution in Eq.~\eqref{eqn_generalevolution_smodes}. Specifically, this involves applying gates. We will delve into gates more in the next section. 

For now, let us remark that gates can be viewed (mathematically) as superoperators that act on the operator space. Superoperator space is the space of linear operators that act on $\mathcal{L}(\mathcal{S})$. We can denote the superoperator space as $\mathcal{L}(\mathcal{L}(\mathcal{S}))$.

In transforming Eq.~\eqref{eqn_barebones_smode} to Eq.~\eqref{eqn_generalevolution_smodes}, we need maps $\mathcal{A}_t$, $B_t$, and $\mathcal{C}_t$ such that
\begin{equation}\label{eqn_superoperators1}
    A(t) = \mathcal{A}_t ( A_0),\quad  \mathbf{b}(t) = B_t(\mathbf{b}_0) ,\quad\text{and}\quad C(t) = \mathcal{C}_t ( C_0)\,.
\end{equation}
Here, $\mathcal{A}_t$ and  $\mathcal{C}_t$ are superoperators and $B_t$ is an operator:
\begin{equation}\label{eqn_superoperators2}
    \mathcal{A}_t , \mathcal{C}_t \in \mathcal{L}(\mathcal{L}(\mathcal{S}))\,,\qquad B_t \in \mathcal{L}(\mathcal{S}) \,.
\end{equation}
The superoperators $\mathcal{A}_t$ and  $\mathcal{C}_t$ map $N\times N$ matrices to $N\times N$ matrices. Hence the superoperator space is $N^4$-dimensional.

For nomenclature purposes, let us call $\mathcal{A}_t$, $B_t$, and $\mathcal{C}_t$ the drift matrix schedule, the drift vector schedule, and the diffusion matrix schedule, respectively. We emphasize that $\mathcal{A}_t$, $B_t$, and $\mathcal{C}_t$ can be viewed as gate sequences, analogous to the gate sequences used to program quantum computers. Importantly, $\mathcal{A}_t$, $B_t$, and $\mathcal{C}_t$ are not pre-encoded into the hardware, but rather they can be viewed as software programs that one runs on the hardware. We elaborate more on these gates and gate sequences in Section~\ref{sc:gates}.

Physically speaking, $\mathcal{C}_t$ can be viewed as a variance schedule for the diffusion process, in that $\mathcal{C}_t$ can serve to either amplify or attenuate the variance associated with the diffusion. Similarly, $\mathcal{A}_t$ and $B_t$ can either accelerate or decelerate the rate of drift.

\subsubsection{Dirac notation}

Following Dirac's work, we can introduce the bra-ket notation for the vector spaces that we are interested in. Using the bra notation, we write the state vector in Eq.~\eqref{eqn_v_1} as
\begin{equation}\label{eqn_braket1}
\mathbf{v}(t) = \sum_{j=1}^N v_j(t)\ket{s_j}
\end{equation} 
where $\{\ket{s_j}\}$ is an orthonormal basis for the state space $\mathcal{S}$.

Similarly, we can expand $A_0$, $\mathbf{b}_0$, and $C_0$ in bra-ket notation as follows:
\begin{equation}\label{eqn_braket2}
A_0 = \sum_{j,k} a_{j,k} \dyad{s_j}{s_k},\qquad \mathbf{b}_0 = \sum_{j=1}^N b_{j}\ket{s_j},\qquad C_0 = \sum_{j,k} c_{j,k} \dyad{s_j}{s_k}\,.
\end{equation} 

Now let us consider the gates that determine the time dependence. The matrix $B_t$ is relative simple and given by
\begin{equation}\label{eqn_braket3}
B_t = \sum_{j,k} b_{j,k} (t)\dyad{s_j}{s_k}\,.
\end{equation} 

For the superoperators $A_t$ and $C_t$, we can use the notation $\ket{s_j,s_k}\rangle = \dyad{s_j}{s_k}$, which is a basis vector for the operator space. The full set $\{\ket{s_j,s_k}\rangle \}_{j,k}$ forms an orthonormal basis for the operator space. Hence we can decompose $A_t$ and $C_t$ in this  orthonormal basis as follows:
\begin{equation}\label{eqn_braket4}
A_t = \sum_{j,k,l,m} a_{jk,lm}(t)\ket{s_j,s_k}\rangle\langle\bra{s_l,s_m},\qquad C_t = \sum_{j,k,l,m} c_{jk,lm}(t)\ket{s_j,s_k}\rangle\langle\bra{s_l,s_m}\,.
\end{equation} 
This concludes our discussion of the vector-space formalism for s-modes.

\subsection{Vector space formalism for s-bits}\label{sc:sbit_formalism}
\subsubsection{State space}
The formalism for s-bits is similar in spirit to that of s-modes. Let us consider a system made of $N$ s-bits, where to each s-bit is associated a state $x \in \{0,1\}$. The state of the full system $\mathbf{x}(t) \in \{0,1\}^N$ can be written as a column vector:
\begin{equation}
    \mathbf{x}(t) = \{x_1, x_2, \ldots, x_N \}^T 
\end{equation}
The $2^N$-dimensional state space $\mathcal{S}^b = \{0,1\}^N$ will be referred to as the s-bit state space, that has the same structure as for a classical spin system.

It is helpful to introduce an additional vector space, on which operators will act. Namely, we let $\mathcal{S}$ denote the vector space $\mathbb{R}^{2^N}$.

\subsubsection{Operator space}

As noted previously, s-bits evolve according to CTMCs. The evolution of the state $\mathbf{x}(t)$ is a sample trajectory from a Markov chain defined by a time-dependent transition matrix $Q(t)$ introduced earlier. A program on N s-bits is a specification of an initial state $\mathbf{x}(0)$ of the s-bits, as well as the rate matrix $Q(t)$ of a desired CTMC. The program directs a collection of physical noise sources that trigger random transitions in the s-bit state stored on digital memory.

The transition matrix $Q(t)$ is called the generator of the Markov chain; a user can generate different Markov chains by altering the time-dependent entries in this matrix. $Q(t)$ belongs to the $4^N$-dimensional space of linear operators $\mathcal{L}(\mathcal{S}^b)$ acting on $\mathbb{R}^{2^N}$. 
\begin{equation}
    Q(t) \in \mathcal{L}(\mathcal{S}^b).
\end{equation}

In practice, one will often be interested in an efficient representation for $Q(t)$. For example, one can consider $Q(t)$ to be a tensor product operator acting independently on each s-bit, or more generally one can consider $Q(t)$ to have some locality in its action.

\subsubsection{Superoperators}

The idea of superoperators acting on s-modes can also be extended to s-bits: a superoperator $\mathcal{Q}_t$ corresponds to a specification on how the transition matrix $Q_0$ can be programmed to be time-dependent:
\begin{equation}
    Q(t) = \mathcal{Q}_t (Q_0).
\end{equation}Therefore, the superoperator $\mathcal{Q}$ belongs to the $16^N$-dimensional linear superoperator space $\mathcal{L}(\mathcal{L}(\mathcal{S}^b))$. We call $\mathcal{Q}_t$ the transition matrix schedule, which specifies how the Markov process transition rules may change in time. Analogously to superoperators in the s-mode case, it can be viewed as a gate sequence used to program the s-bit system. 

Once again, one will typically aim for an efficient representation for $\mathcal{Q}_t$, for example by allowing it to have some locality in its action, such as tensor-product action.

\section{Gates and gate sequences}\label{sc:gates}

\subsection{Continuous approach to gates}

In the following, we restrict the discussion to s-modes for sake of clarity, and also because s-modes have more relevance to the AI applications of interest. Much of the formalism presented hereafter can be extended to s-bits. Specifically, the superoperator $\mathcal{Q}_t$ defined above obeys analogous equations to the ones presented in the following discussion.

We can consider a continuous or discrete approach to gates. Let us first consider a continuous approach. A continuous approach would be analogous to pulse-level control that is employed in quantum computing. Hence we can refer to this approach as pulse-level control.

In the continuous case, we can define the drift and diffusion schedules as continuous functions of time $t$. In this case, the matrix elements of $\mathcal{A}_t$, $B_t$, and $\mathcal{C}_t$ are continuous in $t$. One can therefore write the software program associated with the gate sequence as a description of how the matrix elements of $\mathcal{A}_t$, $B_t$, and $\mathcal{C}_t$ vary with time. The user can directly specify the matrix elements of these matrices. As an example, the user could choose the matrices to be proportional to the identity: \begin{equation}
    \mathcal{A}_t = \alpha(t) I_{N^4},\quad B_t = \beta(t) I_{N^2},\quad \text{and}\quad \mathcal{C}_t = \gamma(t) I_{N^4}
\end{equation} 
and then the user would specify the functions $\alpha(t)$, $\beta(t)$, and $\gamma(t)$.

Alternatively, one can write these matrices in terms of generators, similar to what is done in quantum computing. In quantum computing, one often writes a propagator $U(t_1,t_2)$ as the exponential of a generator $G$, as follows: $U(t_1,t_2) = \exp(-i G (t_2-t_1))$, and in this case $G$ is Hermitian which makes $U$ unitary. For the purpose of gates on Thermodynamic AI systems, we can allow the generator to be more general, either Hermitian or non-Hermitian. Then, we can write the gates as:
\begin{equation}\label{eqn_generator_gates}
    \mathcal{A}_{t_2} = \exp(-i \mathcal{G}_A (t_2-t_1)),\quad B_{t_2} = \exp(-i G_B (t_2-t_1)),\quad \text{and}\quad \mathcal{C}_{t_2} = \exp(-i \mathcal{G}_C (t_2-t_1))
\end{equation} 
where $\mathcal{G}_A $, $G_B $, and $\mathcal{G}_C $ are the user-specified generators for $\mathcal{A}_{t_2}$, $B_{t_2}$, and $\mathcal{C}_{t_2}$ respectively. Here $t_1$ is the starting time of the gate and $t_2$ is the ending time of the gate.

Note that the generator approach is convenient because it avoids having to specify the time-dependence of every matrix element of the gate. Instead, one only needs to specify the generator and the duration $t_2 - t_1$ of the gate.

One can see from Eq.~\eqref{eqn_generator_gates} that the generator approach is reminiscent of pulse-level control. In pulse-level control, one applies a pulse for some user-specified duration. Hence, one can think of the gates in Eq.~\eqref{eqn_generator_gates} as pulses.

\subsection{Discrete approach to gates}

The pulse-level control described above is extremely natural, especially for s-mode systems. We remark that one could take a discrete approach to gates, although it seems less natural. Gates in quantum computing are often denoted as discrete elements, such as X gates or C-NOT gates. However, physically these gates are typically implemented with continuous pulses, and more efficient control can often come from working with continuous pulses rather than with discrete gates. We therefore focus on continuous pulses, even though we also remark that it is possible to form sets of discrete gates out of these continuous pulses.

\subsection{Gate sequence as a software program}

Let us now consider a gate sequence. For our purposes, we will have three gate sequences, one each for $\mathcal{A}_{t}$, $B_{t}$, and $\mathcal{C}_{t}$.  Together, these three gate sequence represent a complete software program. (However, we remark in Section~\ref{sc:entropy_MD} that an additional gate sequence can be employed to program an additional piece of hardware. See Section~\ref{sc:software_MD} for more details.) 

In what follows we adopt the generator-based formalism for writing gate sequences. Suppose that we wish to decompose $\mathcal{A}_{t}$ into a gate sequence of $f$ gates from an initial time $t=t_0$ to a final time $t=t_f$.  Mathematically, we can write this gate sequence as:
\begin{equation}\label{eqn_gatesequence_At}
   \mathcal{A}_t= 
\begin{cases}
   \exp(-i \mathcal{G}_A^{(1)} (t-t_0)),& \text{if } t_0\leq t < t_1\\
    \exp(-i \mathcal{G}_A^{(2)} (t-t_1)),              & \text{if } t_1\leq t < t_2\\
    ...\\
    \exp(-i \mathcal{G}_A^{(f)} (t-t_{f-1})),              & \text{if } t_{f-1}\leq t < t_f
\end{cases}
\end{equation}
Here, we employ a set of generators $\{\mathcal{G}_A^{(1)}, \mathcal{G}_A^{(2)}, ..., \mathcal{G}_A^{(f)}\}$ where $\mathcal{G}_A^{(j)}$ is the generator at the $j$th time interval. Equations analogous to Eq.~\eqref{eqn_gatesequence_At} can also be written for $B_t$ and $\mathcal{C}_t$.

\begin{figure}
    \centering
\includegraphics[width=0.6\textwidth]{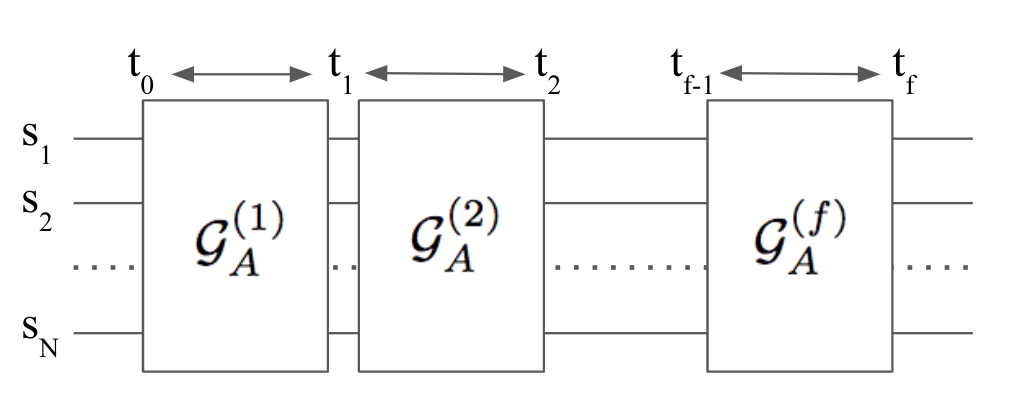}
    \caption{Illustration of a gate sequence associated with the superoperator $\mathcal{A}_t$ given by Eq.~\eqref{eqn_gatesequence_At}. The system consists of $N$ s-modes, and time proceeds from left to right.}
    \label{fig:gatesequence}
\end{figure}

Figure~\ref{fig:gatesequence} provides a diagram of the gate sequence in Eq.~\eqref{eqn_gatesequence_At}. Similar gate sequence diagrams would apply to $B_t$ and $\mathcal{C}_t$. Note that these gate sequences affect the parameters of the dynamics, $A(t)$, $\mathbf{b}(t)$, and $C(t)$, which appear in Eq.~\eqref{eqn_generalevolution_smodes}.

\subsection{Special cases of gates}

\subsubsection{Gates on a single s-mode}

Let us consider the special case where a gate acts only on a single s-mode. This is simplest to describe for the gate $B_t$. Recall that $B_t$ can be written in Dirac notation as $B_t = \sum_{j,k} b_{j,k} (t)\dyad{s_j}{s_k}$. Consider the case where $B_t$ acts non-trivially only on the first s-mode, denoted $s_1$. In this case, we can write:
\begin{equation}\label{eqn_singlesmodegate1}
B_t = b_{1,1}(t)\dyad{s_1}{s_1} + \sum_{j>1} \dyad{s_j}{s_j}
\end{equation} 
Suppose we act with this gate on $\mathbf{b}_0 = \sum_{j=1}^N b_{j}\ket{s_j}$. Then we obtain the resulting vector:
\begin{equation}\label{eqn_singlesmodegate2}
\mathbf{b}(t) = B_t (\mathbf{b}_0) = b_{1,1}(t) b_1 \ket{s_1}+ \sum_{j>1} b_{j}\ket{s_j}
\end{equation} 
One can see that the gate in Eq.~\eqref{eqn_singlesmodegate1} has the effect of multiplying the first element of the $\mathbf{b}_0$ vector by the time-dependent function $b_{1,1}(t)$. Hence, this is a single s-mode gate, acting on the first s-mode. 

Similarly, we can construct a gate $B_t$ that acts on all s-modes independently, without coupling their associated drift vectors. This would be a diagonal matrix, of the form:
\begin{equation}\label{eqn_singlesmodegate3}
B_t =  \sum_{j} b_{j,j}(t)\dyad{s_j}{s_j}
\end{equation} 
The result of applying this gate to $\mathbf{b}_0$ is $\mathbf{b}(t) = \sum_j b_{j,j}(t) b_j \ket{s_j}$. Hence, each component of the drift vector gets multiplied, in an independent fashion, by a time-dependent function. 

Note that we can rewrite this from the perspective of the generator $G_B$ that appears in Eq.~\eqref{eqn_generator_gates}. Namely, the generator associated with the $B_t$ matrix in Eq.~\eqref{eqn_singlesmodegate3} is:
\begin{equation}\label{eqn_singlesmodegate3b}
G_B =  \sum_{j} \beta_j\dyad{s_j}{s_j},\quad\text{with}\quad \beta_j = \frac{i}{\Delta t}\log 
 b_{j,j}(t),
\end{equation} 
where $\Delta t$ is the duration of the pulse.

Constructing single s-mode gates for $\mathcal{A_t}$ and $\mathcal{C_t}$ is more complicated, due to the ambiguous nature of what a single s-mode gate would mean in this case. Nevertheless, it becomes simple for the special case when $A_0$ and $C_0$ are diagonal matrices. In this case, we can write  $A_0 = \sum_{j} a_{j,j} \dyad{s_j}{s_j}$ and $C_0= \sum_{j} c_{j,j} \dyad{s_j}{s_j}$\,. Then, if we want $A_t$ and $C_t$ to act only on the first s-mode, we write:
\begin{align}\label{eqn_singlesmodegate4}
A_t &= a_{11,11}(t)\ket{s_1,s_1}\rangle\langle\bra{s_1,s_1} + \sum_{j>1} \ket{s_j,s_j}\rangle\langle\bra{s_j,s_j}\\
C_t &= c_{11,11}(t)\ket{s_1,s_1}\rangle\langle\bra{s_1,s_1} + \sum_{j>1} \ket{s_j,s_j}\rangle\langle\bra{s_j,s_j}
\end{align} 
Similarly, if we want $\mathcal{A}_t$ and $\mathcal{C}_t$ to act on all s-modes independently (without any coupling between s-modes), then we write:
\begin{align}\label{eqn_singlesmodegate5}
A_t &= \sum_{j} a_{jj,jj}(t)\ket{s_j,s_j}\rangle\langle\bra{s_j,s_j}  \\
C_t &=  \sum_{j} c_{jj,jj}(t) \ket{s_j,s_j}\rangle\langle\bra{s_j,s_j}
\end{align} 

When $A_0$ and $C_0$ have off-diagonal entries, then these matrices describe the connectivity between the s-modes. Hence, at a conceptual level, it is ambiguous as to what a single s-mode gate would be in this case, since such a gate might alter all the connections going into or out of that s-mode, or it might only affect the diagonal elements of $A_0$ and $C_0$. For simplicity, we can consider the latter case, where gate only affects the diagonal elements of $A_0$ and $C_0$ and does not impact their off-diagonal elements. In this case, we write the gates as:
\begin{align}\label{eqn_singlesmodegate6}
A_t &= \sum_{j} a_{jj,jj}(t)\ket{s_j,s_j}\rangle\langle\bra{s_j,s_j}+ \sum_{j,k} \ket{s_j,s_k}\rangle\langle\bra{s_j,s_k}\\
C_t &=  \sum_{j} c_{jj,jj}(t) \ket{s_j,s_j}\rangle\langle\bra{s_j,s_j}+ \sum_{j,k} \ket{s_j,s_k}\rangle\langle\bra{s_j,s_k}
\end{align} 
One can view these gates as acting independently on each individual s-mode and not affecting the coupling between them, since only the diagonal elements of $A_0$ and $C_0$ get multiplied by non-trivial factors.

\subsubsection{Gates that couple multiple s-modes}

In general, gates could couple multiple s-modes together.  In this case, the mathematical forms for the gates correspond to the generic forms already given in Eq.~\eqref{eqn_braket3} and \eqref{eqn_braket4}. Here, the off-diagonal elements in these equations can be non-trivial, which leads to a non-trivial impact on the coupling between s-modes. This concludes our discussion of the different types of gates and gate sequences.

\section{Entropy dynamics, Refrigeration, and Maxwell's Demon}\label{sc:entropy_MD}

\subsection{Innate entropy dynamics of the hardware}

Depending on the choices of the hardware drift  and diffusion matrices, $A_0$ and $C_0$, the entropy of the system may naturally change over time. We call this the innate entropy dynamics of the hardware.

We remark that the form we assumed for the SDE in Eq.~\eqref{eqn_generalevolution_smodes} leads to a Gaussian distribution for the probability distribution at any time $t$, assuming that the initial point $\mathbf{v}(0)$ is a fixed point. Namely, the assumption that the drift term is an affine function of $\mathbf{v}$ leads to the property that resulting probability distribution is Gaussian. Because the probability distributions are Gaussian, then there is a direct relationship between variance and entropy. Hence we can essentially use the terms variance and entropy interchangeably in the following discussion.

For example, one could choose the drift to be zero, $A_0 = 0$ and $\mathbf{b}_0 = \mathbf{0}$. In this case, there is only a diffusion term in the SDE, as follows:
\begin{equation}\label{eqn_VE}
    d\mathbf{v}(t) =  C_0 d\mathbf{w}
\end{equation}
This corresponds to a so-called variance exploding process where both the variance and the entropy grow over time. Hence, for a hardware designed based on Eq.~\eqref{eqn_VE}, the innate entropy dynamics will lead to the entropy monotonically increasing over time.

Alternatively, one could choose:
\begin{equation}\label{eqn_VP}
    A_0 = (-\beta/2) I,\quad \mathbf{b}_0 = \mathbf{0},\quad\text{and}\quad C_0 = \sqrt{\beta} I
\end{equation}
This choice leads to the drift and diffusion balancing each other out, such that the variance (and entropy) decay towards a fixed point of fixed variance (fixed entropy). This is known as a variance preserving process. Hence, for a hardware designed based on Eq.~\eqref{eqn_VP}, the innate entropy dynamics will lead to the entropy settling to a fixed value over time.

While other choices of $A_0$, $\mathbf{b}_0$, and $C_0$ are possible, one can see from the examples in \eqref{eqn_VE} and \eqref{eqn_VP} that the entropy dynamics are often relatively simple, at least for the bare-bones hardware.

\subsection{Complicated entropy dynamics in AI applications}

However, AI applications such as generative modeling or Bayesian inference require complicated entropy dynamics. In generative modeling, one needs to start from a high entropy (i.e., noisy) distribution and gradually reduce the entropy and move the distribution towards a highly structured one that mimics the data distribution. Similarly, in Bayesian inference, the weights of the model must be transformed from a high uncertainty situation (the prior distribution) to a low uncertainty situation (the posterior distribution) as information about the data is introduced during training. 

In both AI applications, the entropy can be viewed as decreasing which might at first seem aphysical, since it sounds contradictory to the second law of thermodynamics. Hence, one may be concerned that one could not implement such complicated entropy dynamics using physics-based hardware. After all, the entropy dynamics discussed above for systems of s-modes are relatively simple, and typically involve the entropy either increasing or staying the same.

Indeed, an isolated system of s-modes will not suffice to achieve the complicated entropy dynamics needed for AI applications. After all, an isolated physical system does obey a local version of the second law, meaning that its entropy cannot decrease over time. 

\subsection{Refrigeration}

On the other hand, an open system (a system that is coupled to an external system) does not obey a local version of the second law. Indeed, this is the basis of refrigeration. Refrigerators locally reduce the entropy of their contents, as the expense of heating up the surrounding air. As a result, global entropy actually increases, since the entropy rise of the air is larger than the entropy decrease of the refrigerator's contents. This suggests that we should bring the system of s-modes in contact with an external system, if we wish for the entropy of the s-mode system to go down over time.

\begin{figure}
    \centering
\includegraphics[width=0.36\textwidth]{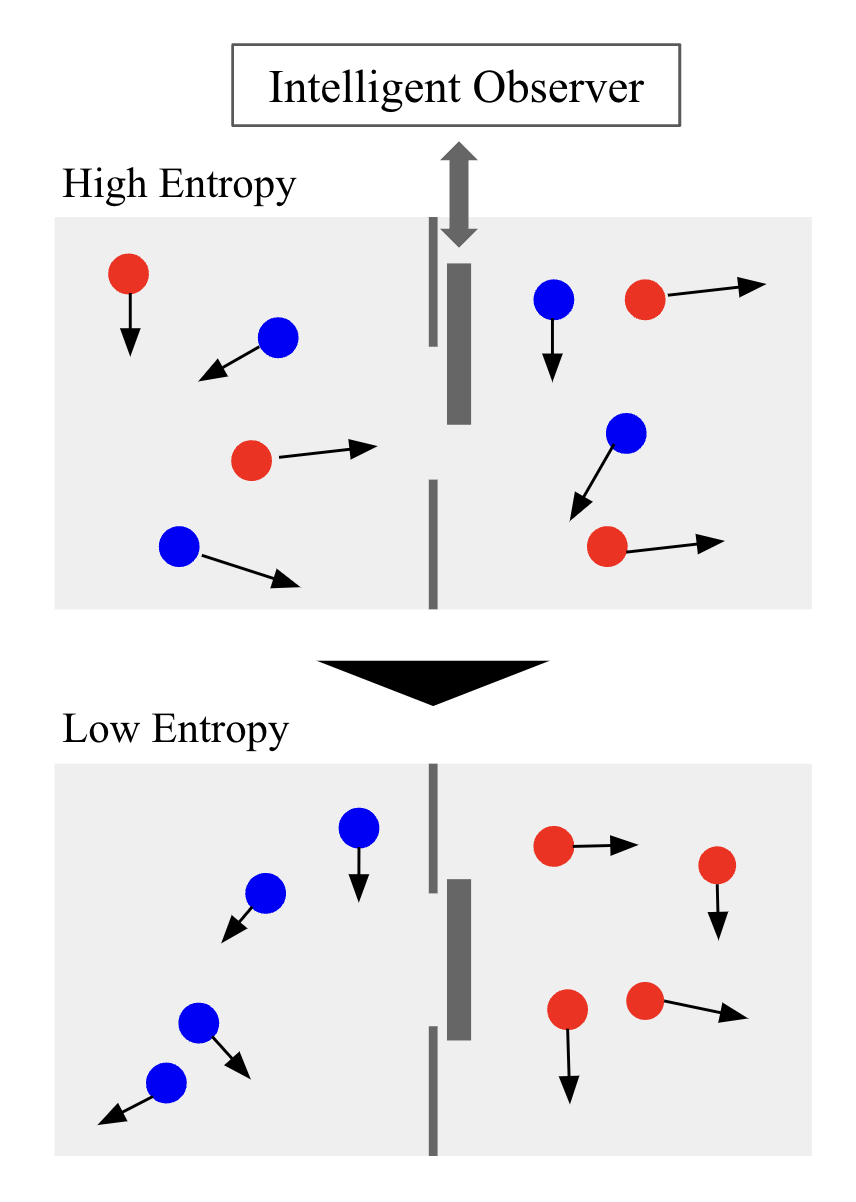}
    \caption{Illustration of Maxwell's Demon. An intelligent observer (the demon) monitors the locations of the components of a gaseous mixture, allowing it to separate the components of the gas, hence reducing the entropy of the system over time.}
    \label{fig:MaxwellDemon_gas}
\end{figure}

\subsection{Maxwell's Demon}

Maxwell's Demon is a thermodynamic concept that is similar to refrigeration, in that it allows a system's entropy to be reduced over time by interacting it with an external system. Here, the demon acts as an intelligent observer who regularly gathers data from (i.e., measures) the system, and based on the gathered information, the demon performs some action on the system. The classic example involves a gaseous mixture and a physical barrier as illustrated in Figure~\ref{fig:MaxwellDemon_gas}. More generally, Maxwell's demon has been experimentally implemented in a variety of platforms~\cite{vidrighin2016photonic,kish2012electrical,cottet2017observing}.

We argue that a Maxwell Demon is both: 
\begin{enumerate}
    \item  Essential to the success of Thermodynamic AI systems due to the complex entropy dynamics required for AI applications, and
    \item Quite straightforward to implement in practice for several different hardware architectures.
\end{enumerate}

Regarding the first point, AI applications like Bayesian inference aim to approximate a posterior distribution, and it is known that such posteriors can be extremely complicated and multi-modal~\cite{izmailov2021bayesian}. Similarly, generative modeling is intended to handle arbitrary data distributions. Hence, producing only Gaussian distributions, as Eq.~\eqref{eqn_generalevolution_smodes} does, will not suffice for these applications.

Regarding the second point, we discuss how one can implement a Maxwell's Demon in hardware in the next subsection.

\subsection{Maxwell's Demon as a hardware component}\label{sc:MD_hardware}

Because the Maxwell's Demon (MD) is an essential part of Thermodynamic AI systems, we regard the MD as being a component of the bare-bones hardware. In addition, one would need a hardware means to communicate or connect the MD hardware component to the hardware system of s-units. One can add a software layer that modifies the communication between the MD hardware and the s-unit hardware. We give more details on this software layer in Section~\ref{sc:software_MD}.

Here, we note that one can construct the MD device in a variety of ways. We consider digital, analog and hybrid digital-analog approaches to building the MD device. Due to their complicated nature, we discuss the analog and hybrid approaches in a later subsection, Sec.~\ref{sc:physical_MD}.

\begin{figure}
    \centering
\includegraphics[width=0.6\textwidth]{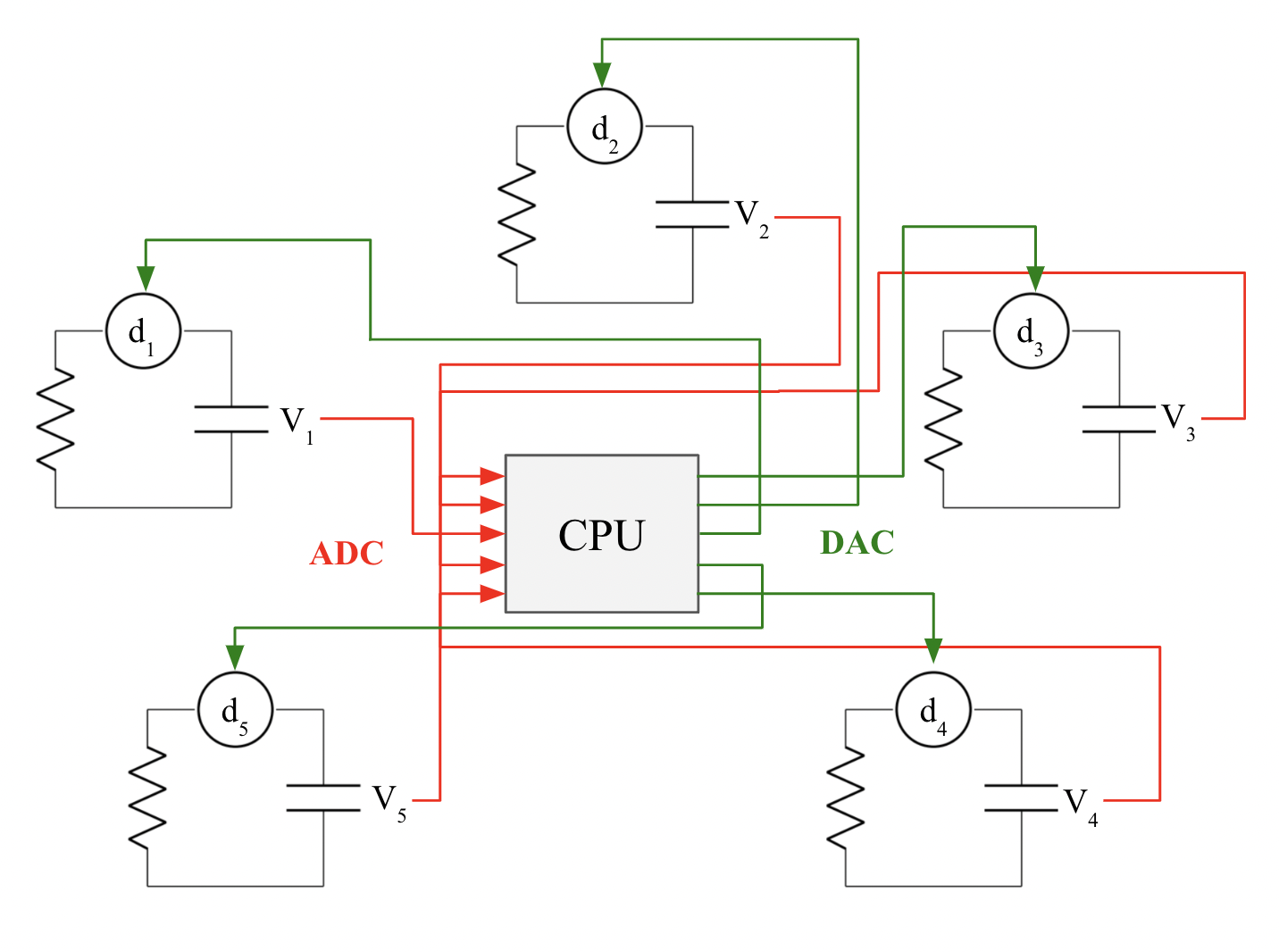}
    \caption{Illustration of how a Maxwell's Demon (stored on a CPU) could interact with a system of s-modes.}
    \label{fig:ScoreNetworkMD}
\end{figure}

On the other hand, a digital approach to the MD system is quite simple to describe. Namely, a Maxwell's Demon could simply correspond to a neural network that is stored on a digital central processing unit (CPU). In this case, one would need to communicate the state vector $\mathbf{v}$ to the CPU (to be the input to the neural network), and then communicate the proposed action of the Maxwell's Demon (i.e., the output of the neural network) back to the thermodynamic hardware. Hence, one simply needs a means to interconvert signals between the thermodynamic hardware and the CPU. This is illustrated in Figure~\ref{fig:ScoreNetworkMD}, with the interconversion shown as analog-to-digital converters (ADCs) and digital-to-analog converters (DACs).

Analog or hybrid approaches to the MD device could allow one to integrate it more closely to the rest of the thermodynamic hardware. Moreover, this could even allow one to avoid having to interconvert signals. Hence it is worth considering such approaches to the MD device, as discussed further in Section~\ref{sc:physical_MD}.

\subsection{Impact of the MD system on the s-unit system}

In principle, one could explicitly model the dynamics of the MD hardware with mathematical equations. Indeed, we delve into the mathematics of the MD hardware in Sec.~\ref{sc:physical_MD}.

Here, we instead focus on the impact of the MD on the dynamics of the s-unit system. Mathematically speaking, the impact of the MD can be viewed as an additional term in the SDE that governs the evolution of the s-units. We describe this as follows.

The key idea is that the MD takes in, as its inputs, the time $t$ and the current value of the state vector $\mathbf{v}(t)$. Because $\mathbf{v}(t)$ is an explicit input to the MD, then this corresponds to the MD monitoring or measuring the state of the s-unit system. This is in analogy to the standard MD paradigm of Fig.~\ref{fig:MaxwellDemon_gas}, where the MD monitors the positions of the gas particles.

The MD then outputs a vector $\mathbf{d}(t, \mathbf{v}(t))$, which we call the demon vector. This vector can be in a variety of physical forms, but it eventually needs to be converted to the physical form that can is acceptable by the s-unit system. Specifically, the demon vector will be applied to the s-unit system is such a way as to give rise to a drift term in the SDE.

The resulting SDE for the bare-bones system of $N$ s-modes then becomes:
\begin{equation}\label{eqn_barebones_smode_MD}
    d\mathbf{v}(t) = (A_0 \mathbf{v}(t)+\mathbf{b}_0+ \mathbf{d}(t, \mathbf{v}(t)))dt + C_0 d\mathbf{w}\,.
\end{equation}
In comparison to Eq.~\eqref{eqn_barebones_smode}, one can see that now the hardware has an additional drift term $\mathbf{d}(t, \mathbf{v}(t))$. This drift term serves to guide the stochastic dynamics in an intelligent way, to achieve the desired goal of the AI task.

\subsection{Training the Maxwell's Demon}\label{sc:MD_training}

Equation~\eqref{eqn_barebones_smode_MD} assumes that the MD already has some level of intelligence, such that the $\mathbf{d}(t, \mathbf{v}(t))$ vector is well suited to solve the AI problem. However, typically one will need to train the MD to be intelligent.

In addition, to achieve complicated probability distributions via the evolution of Eq.~\eqref{eqn_barebones_smode_MD}, the demon vector cannot be some simple function of the input. For example, it cannot be just an affine function of $\mathbf{v}(t)$. Rather, the MD must have a highly expressive, trainable output. So the output should depend on some trainable parameters $\theta$. Training $\theta$ would correspond to teaching the MD to intelligently manipulate the system in order to accomplish the desired AI task.

We write $\mathbf{d}_{\theta}(t, \mathbf{v}(t))$ for the demon vector during the training process, to indicate the dependence on the parameters. Then we write $\mathbf{d}(t, \mathbf{v}(t))$ for the demon vector after training, as in Eq.~\eqref{eqn_barebones_smode_MD}.

\subsubsection{``Ex situ'' training}

In principle, the MD could be trained in isolation from the rest of the Thermodynamic AI hardware. We call this isolated training or \textit{ex situ} training, since it involves training the MD outside of the environment in which it will be used.

This would involving mimicking the s-unit system with, for example, standard digital hardware such as a CPU or field programmable gate array (FPGA). This digital hardware would provide loss function values (or loss function gradient values) for the training process. Here, the loss function is application specific, and could be an evidence lower bound (ELBO) loss function or a score-matching loss function.

After training, one would then hook up the s-unit system to the (pre-trained) MD. There are benefits and drawbacks to this approach. One benefit comes from the fact that the MD might already be stored on a digital device. Then mimicking the s-unit system on that same digital device could allow for an easy training process, without having to convert different types of physical signals. Another benefit is that some industrial applications happen to already have pre-trained neural networks stored on digital devices, and in these cases one can bypass the training step and directly hook up the  s-unit system to the pre-trained MD. 

However, there are some drawbacks to this approach. One drawback is that the physical s-unit system likely has its own physical sources of noise, and these noise sources may be uncharacterized and uncontrollable. In this case, one would not be able to develop a digital model to mimick the s-unit system with full accuracy. Then, after training, when one hooks up the real s-unit system, novel noise sources will appear that were not accounted for in the training process, which could impact the overall performance. An additional drawback is that one cannot take advantage of clever means to estimate loss functions (or their gradients) using the physical s-unit system, since analog physical systems often provide fast shortcuts to estimating quantities that typically appear in loss functions.

\subsubsection{``In situ'' training}\label{sc:insitu}

An alternative approach is to train the MD \textit{in situ}, which means training it in the environment that it will be eventually be used. This involves interacting the (physical) s-unit system with the MD system during the training process.

During the training process, the SDE for the s-unit system is given by:
\begin{equation}\label{eqn_barebones_smode_MD_training}
    d\mathbf{v}(t) = (A_0 \mathbf{v}(t)+\mathbf{b}_0+ \mathbf{d}_{\theta}(t, \mathbf{v}(t)))dt + C_0 d\mathbf{w}\,,
\end{equation}
under the assumption that one is just using the bare-bones hardware, and not running a non-trivial software program. If one is running a non-trivial software program, then Eq.~\eqref{eqn_barebones_smode_MD_training} will get modified by the software program during the training process. We elaborate on the software program during the training process in the next subsection.

One drawback of the \textit{in situ} approach is that, as mentioned above, certain industrial applications already have pre-trained neural networks that can serve as the MD. Hence, \textit{in situ} training cannot be taken advantage of in these cases.

On the other hand, there are several benefits. Analog physical hardware can accelerate the computation of loss functions. For example, computing the norm of a vector and computing a time integral are common subroutines of estimating a loss function, and these can be accelerated by analog physical hardware. Hence, \textit{in situ} training of the MD allows one to use the physical hardware as part of analog strategy to accelerate the estimation of the loss function. 

A second benefit of \textit{in situ} training is learning to correct errors. The idea is to train the MD device to perform well despite being connected to s-unit system that is inherently faulty such that it has uncharacterized and uncontrollable noise sources. If one trains the MD in the presence of these noise sources, then it can learn to correct certain types of imperfections in the s-unit system. After training, the MD will be educated and prepared for these imperfections. In this sense,  \textit{in situ} training makes the MD more robust to noise and errors. We elaborate on this benefit in Section~\ref{sc:error_correction}.

\subsection{Physical approaches to constructing the MD device}\label{sc:physical_MD}

\subsubsection{Considerations when constructing the MD device}

There are several issues to consider when constructing an MD device. For example, one should consider:
\begin{itemize}
    \item expressibility
    \item signal interconversion
    \item latency of communication
\end{itemize}

The issue of expressibility refers to the fact that the MD device typically needs to be a highly expressive model. This means that it should be able to closely approximate a wide variety of target functions. Such expressibility makes the MD device versatile, so that it can be trained to perform a variety of AI tasks on a variety of different datasets. In practice, digital neural network models can offer a higher degree of expressibility than analog neural networks, due to the fact that the former typically have more free parameters than the latter. On the other hand, hybrid digital-analog approaches could still maintain high expressibility, if the digital component has a large number of parameters.

The issue of signal interconversion refers to the fact that additional hardware overhead is needed to convert digital signals to analog ones, and vice versa. The s-mode system of Thermodynamic AI hardware will typically be analog. Hence if the MD system is digital then signal interconversion is necessary (as shown in Fig.~\ref{fig:ScoreNetworkMD}). Employing an analog MD system would allow one to avoid this signal interconversion issue.

Finally there is the issue of latency. There are potentially two sources of latency, the signal interconversion and the MD device itself. To elaborate on the latter source, the MD device may take time to process its input and produce an output (e.g., a forward pass through a neural network involves matrix-vector multiplications and takes some time). Latency can be a significant issue because the s-unit system evolves continuously over time. Its dynamics cannot be paused in order to wait for the output of the MD system. The output of the MD system must be on time, or else the dynamics could be erroneous. 

The issue of latency could be significant for fully digital approaches to the MD system. On the other hand, a fully analog approach to the MD system could eliminate this issue, since no signal interconversion is required and a forward pass through an analog neural network happens essentially instanteously. 

Nevertheless, there is still the expressibility issue with the fully analog approach. Hence a hybrid digital-analog MD device could offer a goldilocks approach. The hybrid approach still maintains high expressibility. In addition, the hybrid approach could either keep the latency low, or if there is some latency, then it may not be a factor. The latter arises because the training process could treat the MD system as a black box that is trained to perform well despite its internal latency. In other words, internal (as opposed to external) latency of the MD system may not be issue because the training process accounts for that.

In what follows, we provide a framework for constructing both fully analog and hybrid digital-analog MD devices.

\subsubsection{Total derivative formalism}

Let us now delve into the idea of physically constructing an MD system. Physical systems typically evolve according to differential equations. Hence this raises the question: Can we describe the output of the MD system in terms of a differential equation? We typically expect the demon vector $\mathbf{d}_{\boldsymbol{\theta}}(t, \mathbf{v}(t)))$ to be a continuous and differentiable function of time $t$. So we can express its time derivative, fairly generally, as follows:
\begin{equation}
    \frac{d}{d t}\big( \mathbf{d}_{\boldsymbol{\theta}}\big) = \mathbf{h}_{\boldsymbol{\theta}}\Big(t,\mathbf{v}, \frac{d\mathbf{v}}{d t},\mathbf{d}_{\boldsymbol{\theta}}\Big)
\end{equation}
where $\mathbf{h}_{\boldsymbol{\theta}}$ is some parameterized function. 

While many possible forms for $\mathbf{h}_{\boldsymbol{\theta}}$ are possible, a particularly elegant form arises from the total derivative formula. The demon vector $\mathbf{d}_{\boldsymbol{\theta}}(t,\mathbf{v}(t))$ is a function of both $t$ and $\mathbf{v}(t)$. Because of this, we can write the total derivative with respect to $t$ as follows:
\begin{equation}\label{eq:score_ODE_3}
    \frac{d}{d t}\big( \mathbf{d}_{\boldsymbol{\theta}}\big) = \frac{\partial \mathbf{d}_{\boldsymbol{\theta}}}{\partial t} 
 + \sum_i \frac{\partial \mathbf{d}_{\boldsymbol{\theta}}}{\partial v_i}  \frac{d v_i}{d\tau}  = \frac{\partial \mathbf{d}_{\boldsymbol{\theta}}}{\partial t} 
 + (\nabla_{ \mathbf{v}} \mathbf{d}_{\boldsymbol{\theta}}) \cdot   \left(\frac{d\mathbf{v}}{d t}\right)
\end{equation}
Here $\nabla_{ \mathbf{v}}$ denotes the gradient with respect to the $\mathbf{v}$ vector. This total derivative expansion provides inspiration for the following model. Let us write the time derivative of the demon vector as:
\begin{equation}\label{eq:score_ODE_4}
    \frac{d}{d t}\big( \mathbf{d}_{\boldsymbol{\theta}}\big) = \mathbf{q}_{\boldsymbol{\theta}}(t, \mathbf{v})
 +  \mathbf{r}_{\boldsymbol{\theta}}(t,\mathbf{v}) \cdot \frac{d\mathbf{v}}{dt} 
\end{equation}
with
\begin{equation}\label{eq:score_ODE_5}
    \mathbf{q}_{\boldsymbol{\theta}}(t,\mathbf{v}) \approx \frac{\partial \mathbf{d}_{\boldsymbol{\theta}}}{\partial t} \quad\text{and}\quad  \mathbf{r}_{\boldsymbol{\theta}}(t,\mathbf{v})\approx \nabla_{\mathbf{v}}\mathbf{d}_{\boldsymbol{\theta}}.  
\end{equation}
In other words, the trainable function $\mathbf{q}_{\boldsymbol{\theta}}(t,\mathbf{v})$ provides a model for the partial derivative $\frac{\partial \mathbf{d}_{\boldsymbol{\theta}}}{\partial t}$, and the trainable function $\mathbf{r}_{\boldsymbol{\theta}}(t,\mathbf{v})$ provides a model for the gradient $\nabla_{\mathbf{v}}\mathbf{d}_{\boldsymbol{\theta}}$. This constitutes a natural splitting of the $\mathbf{h}_{\boldsymbol{\theta}}$ function into two functions $\mathbf{q}_{\boldsymbol{\theta}}$ and $\mathbf{r}_{\boldsymbol{\theta}}$ that describe different contributions to the time derivative of the demon vector.

\subsubsection{Analog and Hybrid MD systems based on total derivative}\label{sc:totalderiv_device}

Equation~\eqref{eq:score_ODE_4} now gives us the flexibility to create either a fully analog MD or a hybrid digital-analog MD.

This MD device will receive analog inputs from the s-mode system, in the form of the vector $\mathbf{v}$ and the time-derivative vector $\frac{d\mathbf{v}}{dt}$. For example, if each s-mode was composed of a resistor and a capacitor, then the voltages across the capacitors would correspond to $\mathbf{v}$ and the voltages across the resistors would correspond to $\frac{d\mathbf{v}}{dt}$.

In the fully analog case, $\mathbf{q}_{\boldsymbol{\theta}}(t,\mathbf{v}) $ and $\mathbf{r}_{\boldsymbol{\theta}}(t,\mathbf{v}) $ are the outputs of analog neural networks. In the hybrid digital-analog case, these quantities are the outputs of neural networks that are stored and processed on a digital device, such as an FPGA, and ADCs and DACs are employed for interconversion of analog and digital signals. Regardless, we imagine that eventually the $\mathbf{q}_{\boldsymbol{\theta}}(t,\mathbf{v}) $ and $\mathbf{r}_{\boldsymbol{\theta}}(t,\mathbf{v}) $ outputs are provided as analog signals. Then one can construct analog methods to perform the multiplication and addition operations needed to compute the right-hand-side of Eq.~\eqref{eq:score_ODE_4}, and finally one can analog time integrate the differential equation in Eq.~\eqref{eq:score_ODE_4} to produce the demon vector.

\subsubsection{Alternative MD system based on analog forces}\label{sc:analogforces}

\begin{figure}
    \centering
\includegraphics[width=0.5\textwidth]{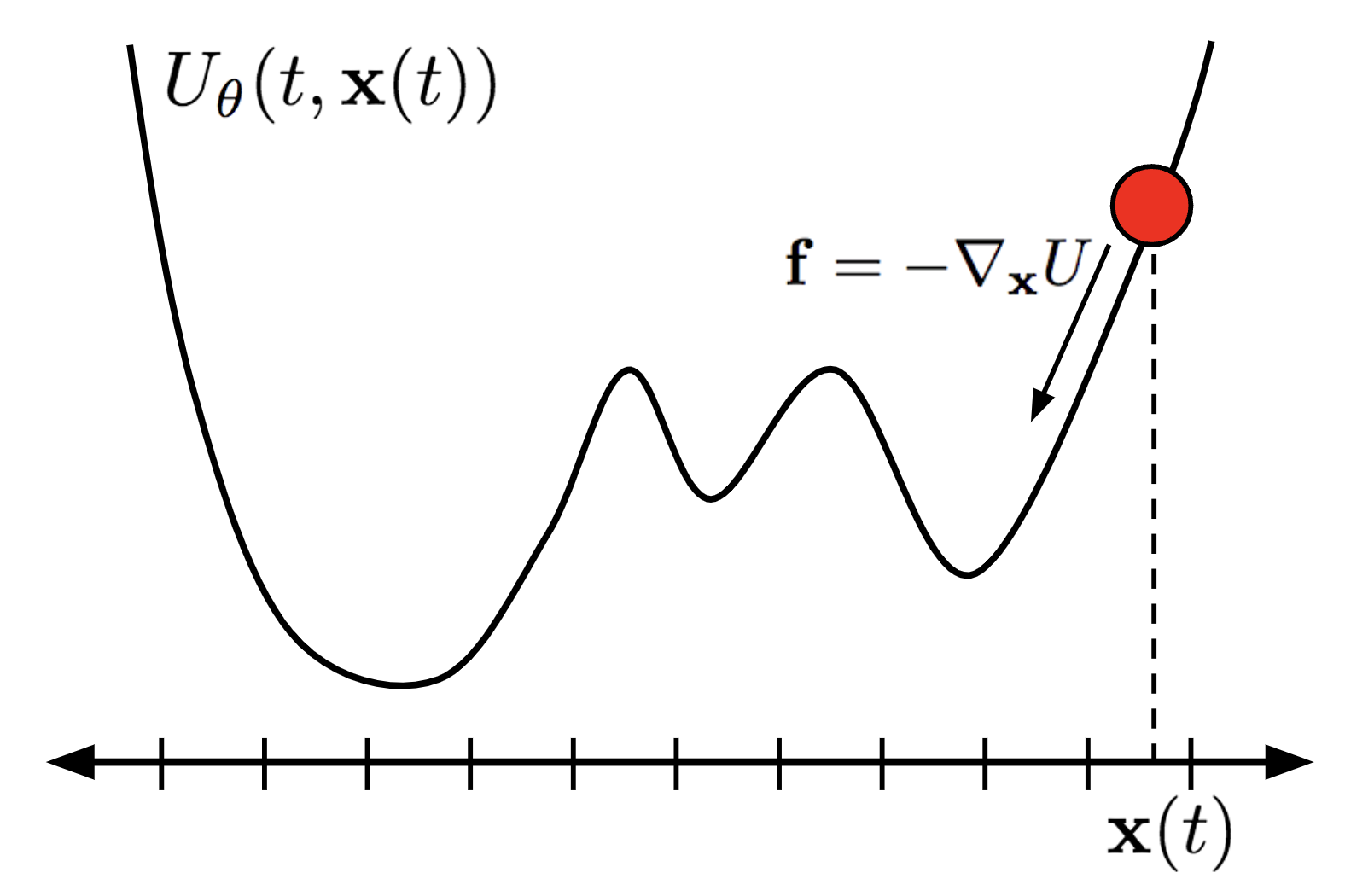}
    \caption{Force-based approach to constructing a Maxwell's Demon device.}
    \label{fig:Potential}
\end{figure}

Let us discuss an alternative approach to constructing the MD system. The following approach will be especially relevant to the applications of Monte Carlo inference and annealing-based optimization.

Ideally we would like an approach that lends itself well to analog, physical implementation. One of the first concepts that comes to mind when thinking of physics is forces. Our strategy will be to think of the output of the MD system as a force. After all, the force is a vector quantity, just like the demon vector.

Recall from Newtonian mechanics that the force $\mathbf{f}(\mathbf{x})$ on a system at position $\mathbf{x}$ is given by the gradient of the potential energy function, $U(\mathbf{x})$,
\begin{equation}\label{eq:force1}
 \mathbf{f}(\mathbf{x}) =  - \nabla_{\mathbf{x}} U(\mathbf{x})\,.
\end{equation}
In addition, the time derivative of the position is given by the momentum vector $\mathbf{p}$,
\begin{equation}\label{eq:force2}
 \frac{d\mathbf{x}}{dt} =  - M^{-1}\mathbf{p}\,.
\end{equation}
Here $M$ is the mass matrix.

Equations~\eqref{eq:force1} and \eqref{eq:force2} provide a recipe for how to construct an MD device. The idea would be to view the  momentum $\mathbf{p}$ as the state vector associated with the s-mode system. Hence, the momentum will evolve over time according to the SDE equation associated with the s-mode system, such as Eq.~\eqref{eqn_generalevolution_smodes}. At a given time $t$, the MD system will take the momentum vector in as an input. Then the MD system will output a force that is a function of both the time $t$ and the momentum $\mathbf{p}(t)$,
\begin{equation}\label{eq:force3}
\text{Input from s-modes: }\mathbf{p}(t),\qquad\text{Output to s-modes: }\mathbf{f}(t,\mathbf{p}(t)) 
\end{equation}

Let us discuss how the MD device performs the mapping from input to output in Eq.~\eqref{eq:force3}. The idea is that the MD device has a latent variable or hidden variable, which corresponds to the position vector $\mathbf{x}(t)$. The latent variable $\mathbf{x}(t)$ is stored inside the MD device's memory, and it evolves over time. Specifically, it evolves over time according to the differential equation in Eq.~\eqref{eq:force2}. Hence, for the latent variable, we can write:
\begin{equation}\label{eq:force4}
 \text{Latent variable:}\quad\mathbf{x}(t) = \mathbf{x}(0) + \int_{t'=0}^{t'=t} \frac{d\mathbf{x}}{dt'} dt' =  \mathbf{x}(0) - \int_{t'=0}^{t'=t} M^{-1}\mathbf{p}(t')dt'\,,
\end{equation}
where $\mathbf{x}(0)$ is some initial starting point for the latent variable.

The MD device also stores a potential energy function $U_{\theta}(t, \mathbf{x}(t))$. For generality, we allow this potential energy function to be time-dependent. This time dependence is important for certain applications such as annealing, where one wishes to vary the potential energy function over time. In addition, we also allow the potential energy function to depend on a set of parameters $\theta$. These parameters are trainable and will be trained during the training process, as discussed in Sec.~\ref{sc:MD_training}. Hence, the potential energy function represents the trainable portion of the MD device,
\begin{equation}\label{eq:force5}
 \text{Trainable Potential Energy Function:}\quad U_{\theta}(t, \mathbf{x}(t))\,.
\end{equation}

Internally, the MD device combines Equations~\eqref{eq:force4} and \eqref{eq:force5} in order to produce a force. Specifically, the MD device employs Eq.~\eqref{eq:force1} and computes the gradient of the potential energy function at time $t$ and location $\mathbf{x}(t)$. This is illustrated in Figure~\ref{fig:Potential}. The result of this computation essentially corresponds to the demon vector $\mathbf{d}_{\theta}$. Hence, the output of the MD device is given by:
\begin{equation}\label{eq:force6}
 \text{Output:}\quad \mathbf{d}_{\theta}(t,\mathbf{v}(t)) = \mathbf{d}_{\theta}(t,\mathbf{p}(t))=  - \nabla_{\mathbf{x}} U_{\theta}(t, \mathbf{x}(t))\,.
\end{equation}
Here we note that we employed the notation $\mathbf{p}(t)$ for the state variable of the s-mode system, although in all other sections of this paper we used $\mathbf{v}(t)$ for this state variable. This concludes our discussion of the mathematical formalism needed for force-based MD devices. 

\subsubsection{Physical implementation of force-based MD device}

We can now remark on how this formalism lends itself nicely to analog, physical implementation. 

Forces are the fundamental ingredients in classical Newtonian physics. Hence, the dynamics of all classical systems can be described by the forces that act on upon them. This includes mechanical systems, electrical systems, optical systems, and magnetic systems. The basic ingredients to constructing a force-based MD device is to have a classical physical system with a continuous state variable, such as position in mechanical systems, charge or current in electrical systems, or electric field or magnetic field in optical systems.

There are two key computations required to output a force:
\begin{itemize}
    \item Evolving the latent variable over time according to Eq.~\eqref{eq:force2}.
    \item Computing the gradient of the potential energy function as in Eq.~\eqref{eq:force1}.
\end{itemize}
In principle, each of these two subroutines could be performed on standard, digital hardware. Indeed, this would offer a unique perspective for certain applications, e.g., it would provide a novel means of constructing (digital) score networks for diffusion models.

\begin{figure}
    \centering
\includegraphics[width=0.4\textwidth]{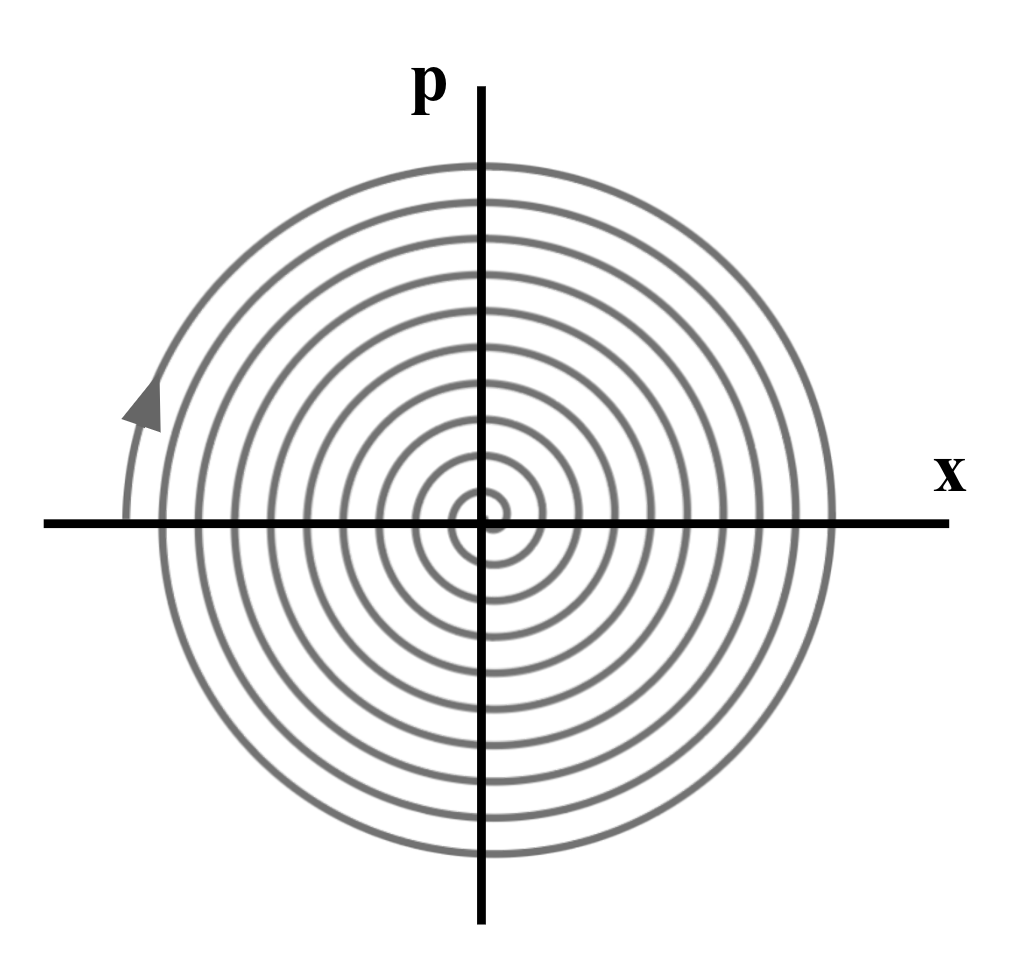}
    \caption{Example of phase space dynamics for a damped physical system.}
    \label{fig:phasespace}
\end{figure}

However, the key idea is that both of these subroutines are amenable to be accelerated by physical processes on analog hardware. One can also mix and match, in the sense of analog hardware can be used to accelerate one of the subroutines, while the other subroutine is performed digitally.

Let us consider the subroutine of evolving the latent variable. With physics-based hardware, this subroutine happens very naturally. Specifically, this essentially corresponds to phase space dynamics associated with Hamilton's equations. For illustration we show an example of phase space dynamics for a damped system in Fig.~\ref{fig:phasespace}. The idea is that phase space dynamics happen naturally in physical systems, and hence we can leverage this.

Specifically, suppose that the continuous state variable $\mathbf{v}$ for the s-mode system corresponds to some physical dynamical variable. Then the key idea is to encode the latent variable $\mathbf{x}$ in the conjugate variable of $\mathbf{v}$. In physics, dynamical variables come in pairs: position and momentum are conjugate variables in mechanical systems, charge and current are a conjugate variables in electrical systems. By encoding the latent variable of the MD device in the conjugate variable of $\mathbf{v}$, then we ensure that the two variables obey a form of Hamilton's equations, and hence they evolve together in phase space with the desired relation.

Let us now consider the other subroutine of computing the gradient of the potential energy function. One straightforward way of doing this is employing a digital neural network whose output is the potential energy function $U_{\theta}$, and then digitally computing the gradient. However, this does not take advantage of physics. 

An alternative means of doing this would be to have a parameterized physical system whose physical potential energy function corresponds to $U_{\theta}$. With this approach, the gradient $\nabla U_{\theta}$ corresponds to the force on the physical system. Hence, one could simply measure this force (as opposed to computing it), and then output this meausured force as the demon vector. This avoids having to compute the gradient of the potential energy function.

\subsection{Software-level programming of the Maxwell's Demon}\label{sc:software_MD}

\subsubsection{Programming the the coupling between the MD and the s-units}

The demon vector appear in Eq.~\eqref{eqn_barebones_smode_MD} opens up another avenue for software-level programming. We introduce the matrix $D_0$, which acts on the demon vector in the SDE associated with the bare-bones hardware, as follows:
\begin{equation}\label{eqn_barebones_smode_MD_D0}
    d\mathbf{v}(t) = (A_0 \mathbf{v}(t)+\mathbf{b}_0+ D_0\mathbf{d}(t, \mathbf{v}(t)))dt + C_0 d\mathbf{w}\,.
\end{equation}

Analogous to the discussion in Section~\ref{sc:superoperator}, we can introduce a superoperator that acts on $D_0$, in order to make it time dependent. Namely, we write:
\begin{equation}\label{eqn_superoperators_Dt}
    D(t) = \mathcal{D}_t ( D_0)
\end{equation}
Here, $\mathcal{D}_t$ is the superoperator that acts as a gate or gate sequence, and it transforms $D_0$ to $D(t)$. 

Hence, $\mathcal{D}_t$ can be viewed as something that the user chooses at the software level. Conceptually, choosing $\mathcal{D}_t$ corresponds to choosing the coupling between the MD device and the s-mode device.

When all superoperators are applied (i.e., the superators in both Eq.~\eqref{eqn_superoperators_Dt} and Eq.~\eqref{eqn_superoperators2}), then the resulting SDE is:
\begin{equation}\label{eqn_fullyprogrammed_SDE}
    d\mathbf{v}(t) = (A(t) \mathbf{v}(t)+\mathbf{b}(t)+ D(t)\mathbf{d}(t, \mathbf{v}(t)))dt + C(t) d\mathbf{w}\,.
\end{equation}
This corresponds to the dynamics of the s-mode device when the user has uploaded their software program to the Thermodynamic AI hardware.

\subsubsection{Programming the MD itself}\label{sc:programmingMD}

Finally, we note that one could program the Maxwell's Demon itself. The nature of this programming depends highly on the physical realization of the MD device. In particular, one has the option of either:
\begin{enumerate}
    \item Storing and processing the MD system on digital hardware, such as a CPU or FPGA, as noted in Sec.~\ref{sc:MD_hardware}.
    \item Building a analog or hybrid digital-analog version of a MD device, as discussed in Sec.~\ref{sc:physical_MD}. 
\end{enumerate}
 
If one employs the first option, then all of the standard software tools used to program neural networks will be available to the user, to program the MD system. 

Similarly, if one employs a hybrid digital-analog approach, as discussed in Sec.~\ref{sc:physical_MD}, then all of the standard software tools used to program neural networks will be available to the user, in order to program the neural networks whose outputs are $\mathbf{q}_{\boldsymbol{\theta}}(t,\mathbf{v}) $ and $\mathbf{r}_{\boldsymbol{\theta}}(t,\mathbf{v}) $.

If one employs a fully analog approach, then the user will likely have less flexibility to alter the basic structure of the (analog) neural network. Nevertheless, the user can still choose the optimization algorithm for the training process.

\section{Thermodynamic Error Correction and Noise Robustness}\label{sc:error_correction}

\subsection{Noise plaguing other computing paradigms}

It is no secret that hardware noise has plagued particular computing paradigms. As we highlighted in the Introduction, noise is the key issue that is preventing quantum computing from being a commercially relevant technology. Noise in quantum computing can turn otherwise efficient algorithms (with polynomial scaling) into inefficient algorithms (with exponential scaling). This essentially destroys the so-called quantum speedup that one would hope for over classical methods. 

Analog computing is another paradigm that suffers from hardware noise. Analog computing has a rich history, and at one point it was perhaps the dominant form of computing. However, digital computers become more precise and more economical in the 1950s to 1970s, which led to the decline of analog computing. Hardware noise in analog computing also played a role in their decline relative to digital computers.

\subsection{Using noise to one's advantage}

In contrast, in Thermodynamic AI, noise is part of the design. It is a fundamental ingredient in the hardware. While quantum and analog computing view noise as a nuisance, Thermodynamic AI views noise as essential. For example, one can see Section~\ref{sc:stochasticity}, where we made a detailed argument for the utility of noise in AI applications. Thus, Thermodynamic AI turns lemons into lemonade. 

While the precise physical construction of Thermodynamic AI hardware is flexible and up for debate, it is worth noting that a likely candidate would be using the same kind of analog components used in analog computing.  These analog circuits will be noisy, of course. But this (unintentional) noise can be essentially combined with whatever noise sources that one is intentionally engineering into the Thermodynamic AI hardware. The final noise that results with thus be a combination of unintentional and intentional noise.  We elaborate below on how this is essentially becomes a non-issue, especially when a Maxwell's Demon is involved in the Thermodynamic AI system.

\subsection{Noise preserves the mathematical framework}

Suppose that one intentionally designs the hardware to have a drift matrix, drift vector, and diffusion matrix given respectively by $A_0$, $\mathbf{b}_0$, and $C_0$. Then naively one might thing that the s-mode system would evolve according to Eq.~\eqref{eqn_barebones_smode}, repeated here for convenience:
\begin{equation}\label{eqn_barebones_smode_again}
    d\mathbf{v}(t) = (A_0 \mathbf{v}(t)+\mathbf{b}_0)dt + C_0 d\mathbf{w}\,.
\end{equation}

However, in practice there will be unintentional and uncharacterized hardware noise. An example of unintentional noise would be the capacitances of capacitors, or resistances of resistors, being off from their nominal values. Another example would be having an additional source of stochastic noise leading to an additional diffusion term.

Mathematically, we encompass these hardware noise sources with a simple model. Suppose that the true values of the relevant matrices and vectors are perturbed away from $A_0$, $\mathbf{b}_0$, and $C_0$. Let us write this perturbation as:
\begin{equation}\label{eqn_barebones_smode_perturbed}
    \tilde{A}_0 = A_0 + A_e,\qquad 
    \tilde{\mathbf{b}}_0 = \mathbf{b}_0 + \mathbf{b}_e,\qquad  \tilde{C}_0 = C_0 + C_e\,.
\end{equation}
Here, $A_e$, $\mathbf{b}_e$, and $C_e$ are the errors or perturbations, and the true objects implemented in hardware are $\tilde{A}_0$, $\tilde{\mathbf{b}}_0$, and $\tilde{C}_0$. Hence, the true evolution of the s-mode system is given by:
\begin{equation}\label{eqn_barebones_smode_effective}
    d\mathbf{v}(t) = ( \tilde{A}_0 \mathbf{v}(t)+\tilde{\mathbf{b}}_0)dt + \tilde{C}_0  d\mathbf{w}\,.
\end{equation}

Equation~\eqref{eqn_barebones_smode_effective} represents the effective design of the bare-bones hardware. It has the same mathematical form as Eq.~\eqref{eqn_barebones_smode_again}, but with different hyperparameters. From a mathematical perpective, it is as if the hardware designer intentionally designed the hardware with the hyperparameters $\tilde{A}_0$, $\tilde{\mathbf{b}}_0$, and $\tilde{C}_0$. Hence, Eq.~\eqref{eqn_barebones_smode_effective} still fits within the mathematical framework of Thermodyanmic AI hardware.

We remark that one could have a multiplicative perturbation instead of the additive perturbation in Eq.~\eqref{eqn_barebones_smode_perturbed}, and the noise robustness argument discussed above would still hold.

\subsection{Maxwell's Demon learns to correct errors}

In Section~\ref{sc:entropy_MD}, we argued that a Maxwell's Demon (MD) device was a key ingredient in Thermodynamic AI hardware, since it allows one to produce complicated entropy dynamics and complicated probability distributions. It turns out there is yet another benefit of employing a MD in the Thermodynamic AI hardware: error correction.

Let us consider the noise model introduced above in Eq.~\eqref{eqn_barebones_smode_perturbed}. One might be worried that noise perturbing the hyperparameters would affect the overall performance of the Thermodynamic AI hardware. But that is where the MD system comes to the rescue.

Suppose that we perform \textit{in situ} training of the MD, as described in Section~\ref{sc:insitu}. This means that the MD is trained in the presence of the physical s-mode system. During training, one employs a loss function that is an objective, true measure of the performance on the AI task.  This loss function treats the hardware as a black box that produces outputs - the loss is agnostic to the internal workings of the hardware. Hence, if one is able to successfully minimize the loss function, then this implies that the system truly is performing well. It also implies that the MD system is well-suited to guide the hardware system to accomplish the desired AI task, regardless of the hardware's noise.

Hence, let us make the assumption that one is able to successfully minimize the loss function during training. Furthermore, let us make the assumption that whatever noise processes that are present during training are also the same  noise processes after training. In other words, let us assume that the perturbations in Eq.~\eqref{eqn_barebones_smode_perturbed} do not change significantly over time, such that whatever noise is present during training is also present after training.

Under these assumptions, the trained MD system will automatically correct for errors or noise in the hardware. Training in the presence of noise makes the system robust to that same noise. Essentially, the MD system learns to account for the hardware noise whenever it produces its outputs, so that the demon vector $\mathbf{d}(t, \mathbf{v}(t))$ will be appropriately adjusted to deal with the hardware noise.

In this sense, Thermodynamic AI systems have inherent robustness to hardware noise.

\section{Application: Thermodynamic Diffusion Models}\label{sc:diffusionmodels}

\begin{figure}
    \centering
\includegraphics[width=0.8\textwidth]{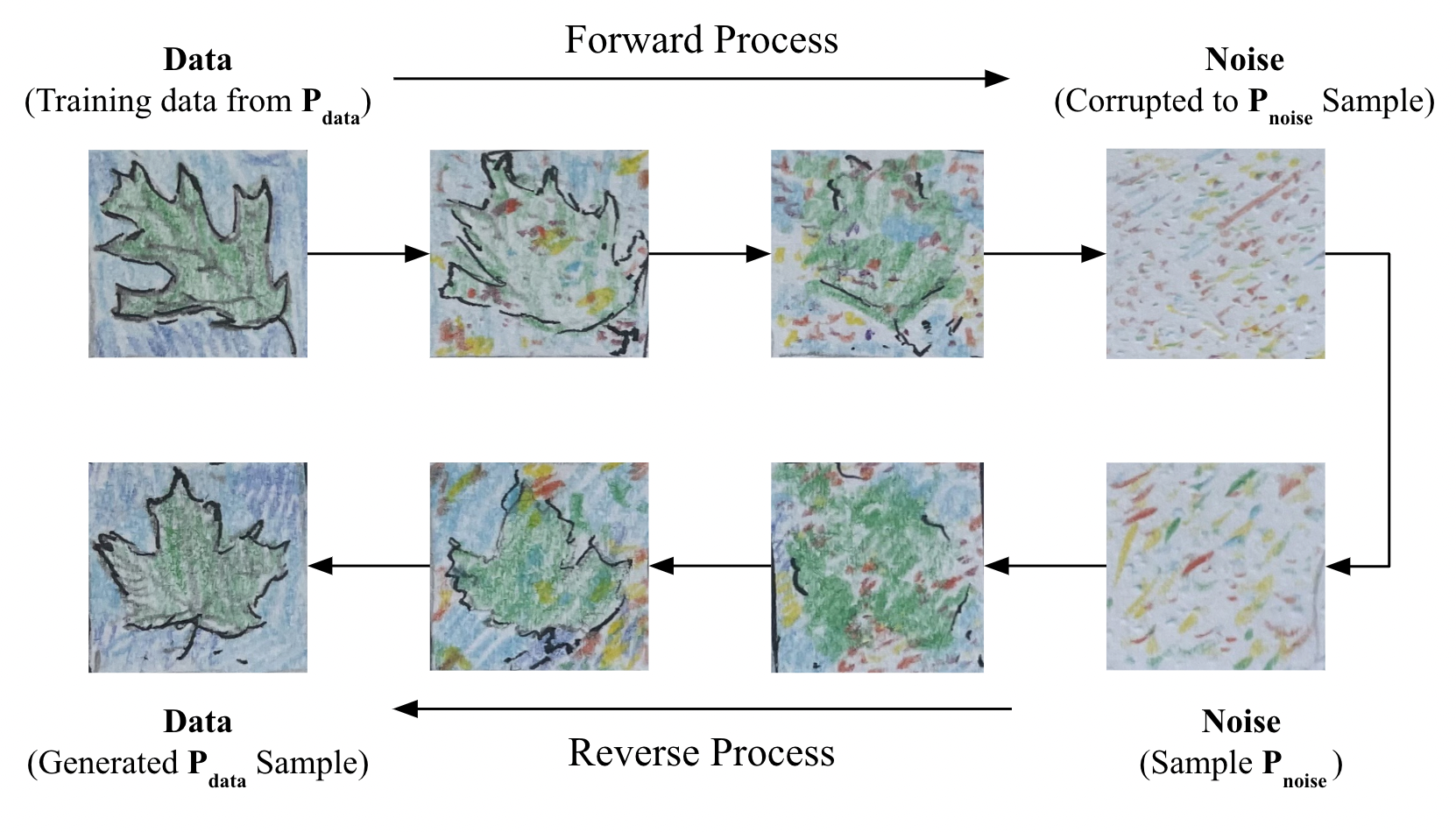}
    \caption{Diffusion models add noise to data drawn from a data distribution $p_{data}$ during a forward process. This provides training data to train a score network. A sample is then drawn from a noisy distribution, $p_{noise}$, and the process is reversed to generate a novel datapoint, with the trained score network guiding the reverse process.}
    \label{fig:DiffusionProcess}
\end{figure}

\subsection{Background}

Diffusion models~\cite{ho2020denoising,song2020score} are a state-of-the-art method for implementing generative models. They have been applied to protein design~\cite{watson2022broadly}, drug design~\cite{schneuing2022structure}, and solid-state crystal design~\cite{xie2021crystal}. Moreover, they were the underlying model responsible for text-to-image technology employed in Stable Diffusion.

Figure~\ref{fig:DiffusionProcess} shows the basic idea of diffusion models.  These models add noise to data drawn from a data distribution $p_{data}$ during a forward process that evolves from from $t=0$ to $t=T$. A sample is then drawn from a noisy distribution, $p_{noise}$, and the process evolves in the reverse direction, from $t=T$ to $t=0$, to generate a novel datapoint.

Both the forward and reverse processes can be described by SDEs~\cite{song2020score}. The forward SDE typically has a mathematical structure corresponding to that of 
Eq.~\eqref{eqn_generalevolution_smodes}. On the other hand, the reverse SDE has a structure corresponding to that of Eq.~\eqref{eqn_fullyprogrammed_SDE}. 

Namely, the reverse SDE is similar to the forward SDE except that it has an additional drift term. A typical forward and reverse SDE for diffusion models has the form:
\begin{align}\label{eqn:VPVEunified1}
        \mathrm{d}\mathbf{x} &= f(t)\mathbf{x} \mathrm{d}t + g(t) \mathrm{d}\mathbf{w}_{t} \quad \text{(Forward)}\\
        \label{eqn:VPVEunified2}\mathrm{d}\mathbf{x} &= f(t)\mathbf{x} \mathrm{d}t - g(t)^2 \mathbf{s}_\theta(\mathbf{x},t) \mathrm{d}t + g(t)\mathrm{d}\mathbf{\overline{w}}_{t} \quad \text{(Reverse)}
\end{align}
where $\mathbf{x}$ is the continuous state variable. In the reverse SDE, time $t$ runs backwards and $\mathrm{d}\mathbf{\overline{w}}_{t}$ is a standard Brownian motion term when time runs backwards. The vector $\mathbf{s}_\theta(\mathbf{x},t)$ is a model for the score function $\nabla_{\mathbf{x}} \log p_t(\mathbf{x})$ (the gradient of the logarithm of the probability distribution), and hence
$\mathbf{s}_\theta(\mathbf{x},t)\approx \nabla_{\mathbf{x}} \log p_t(\mathbf{x})$.

At a high level, the forward process provides training data to train a neural network, whose job is to output $\mathbf{s}_\theta(\mathbf{x},t)$. This neural network is called the score network. Once the score network is trained, it can be used to guide the reverse process to generate novel samples.

In state-of-the-art diffusion models, the sampling rate still remains fairly slow. Hence, any means to speed up sampling rate could significantly improve this technology, and this is where Thermodynamic AI Hardware could help.

\subsection{Fitting into our Thermodynamic AI framework}

Equations~\eqref{eqn:VPVEunified1} and \eqref{eqn:VPVEunified2} respectively fall under the framework of our Eq.~\eqref{eqn_generalevolution_smodes} and Eq.~\eqref{eqn_fullyprogrammed_SDE}. To make this more clear, one can do a change of variables $\tau = T-t$ in the reverse process and rewrite the diffusion model SDE equations as:
\begin{align}\label{eqn:VPVEunified1_new}
        \mathrm{d}\mathbf{x} &= f(t)\mathbf{x} \mathrm{d}t + g(t) \mathrm{d}\mathbf{w}_{t} \quad \text{(Forward)}\\
        \label{eqn:VPVEunified2_new}\mathrm{d}\mathbf{x} &= - f(T-\tau)\mathbf{x} \mathrm{d}\tau + g(T-\tau)^2 \mathbf{s}_\theta(\mathbf{x},T-\tau) \mathrm{d}\tau + g(T-\tau)\mathrm{d}\mathbf{w}_{\tau} \quad \text{(Reverse)}
\end{align}
In this case, both $t$ and $\tau$ run forward in time, i.e., $dt$ and $d\tau$ are positive increments in these equations. Note that $\mathrm{d}\mathbf{w}_{\tau}$ appears in this equation since $\mathrm{d}\mathbf{w}_{\tau} = \mathrm{d}\mathbf{\overline{w}}_{t} $.

Since the time variables in these equations run forward in time, then they can be directly implemented in physical analog hardware, since time always runs forward in the physical world.

The key to fitting diffusion models into our framework is to make to following mapping:
\begin{align}
    \text{(diffusion process)} &\leftrightarrow  \text{(s-mode device)}\\
    \text{(score network)} &\leftrightarrow  \text{(Maxwell's demon device)}
\end{align}
The mathematical diffusion process in diffusion models can be mapped to the physical diffusion process in the s-mode device. Similarly, the score vector outputted by the score network corresponds to the demon vector outputted by the MD device. Hence, diffusion models fit in our framework for Thermodynamic AI systems.

\begin{figure}
    \centering
\includegraphics[width=0.9\textwidth]{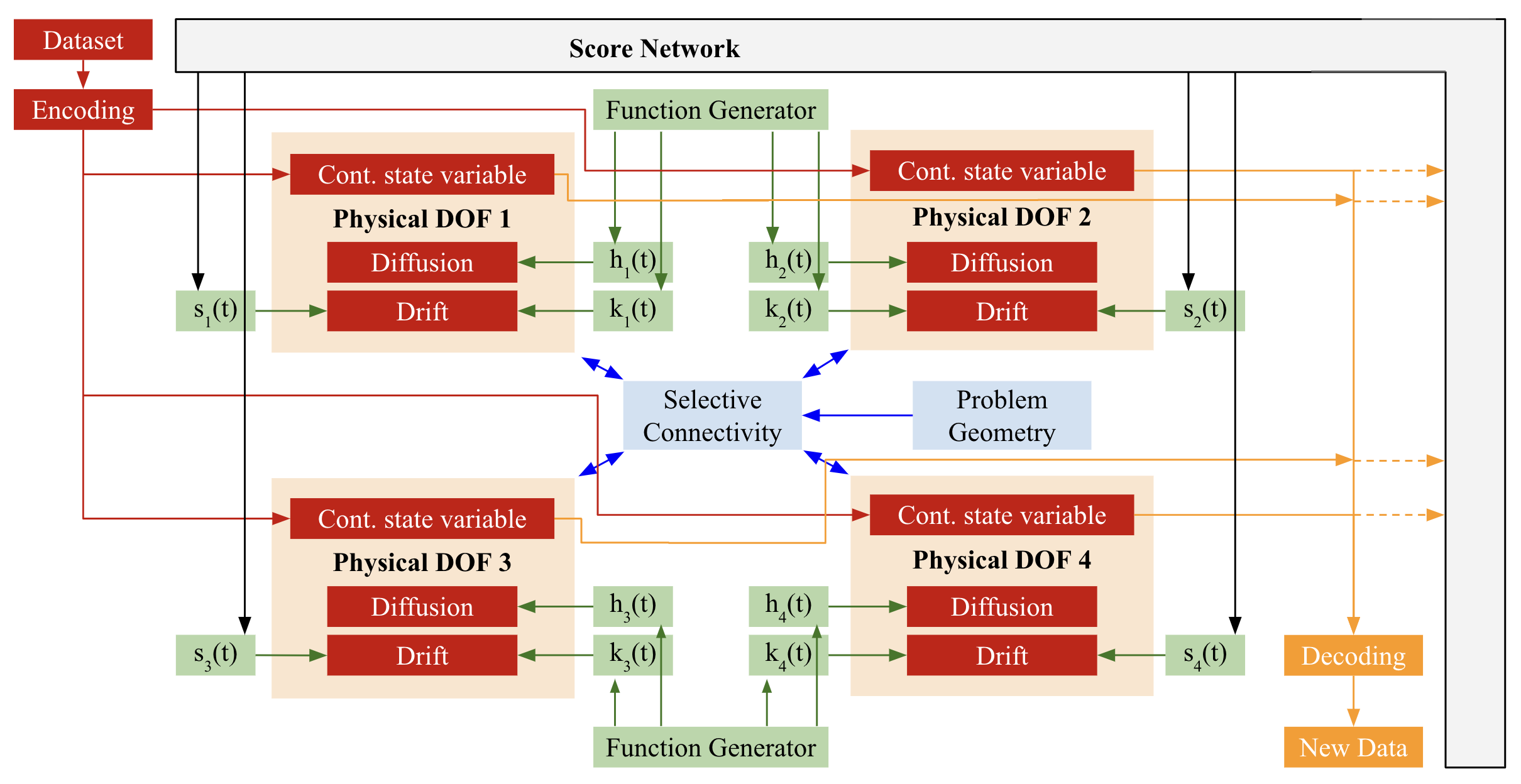}
    \caption{Schematic diagram of a Thermodynamic Diffusion Model. For simplicity, we show the case of four degrees of freedom (DOFs), where each DOF corresponds to an s-mode. These s-modes can be constructed as described in Sec.~\ref{sc:phys_smode}.}
    \label{fig:GenericReverse}
\end{figure}

\subsection{Description of Diffusion Hardware}

Figure~\ref{fig:GenericReverse} gives a schematic diagram of a Thermodynamic Diffusion Model. A variety of different physical paradigms can be used to implement this hardware, such as analog electrical circuits or continuous-variable optical systems. Hence, we describe the system an abstract level.

As shown in Figure~\ref{fig:GenericReverse} the physical system has multiple degrees of freedom (DOFs) - which essentially correspond to the s-modes. The number of DOFs matches the dimensionality of the data, i.e., the number of features in the data. Each DOF has a continuous state variable, and that variable evolves according to a differential equation that, in general, could have both a diffusion and drift term. A function generator can multiply these diffusion and drift terms by arbitrary time-dependent functions ($h_j(t)$ and $k_j(t)$ respectively). The problem geometry, associated with a given dataset, can be uploaded onto the device by selectively connecting the various DOFs, which mathematically couples the differential equations of the various DOFs. After some encoding, a datapoint from the dataset of interest can be uploaded to the device by initializing the values of the continuous state variables to be the corresponding feature values of the datapoint. Similarly, data can be downloaded (and decoded) from the device by measuring the values of the continuous state variables after some time evolution.

In addition, the reverse process uses a trained score network. The inputs to the score network are the values of the continuous state variables at some time $t$, and the output is the value of the score. The $j$th component of the score, $s_j(t)$, gets added as a drift term in the evolution of the $j$th DOF. The score network acts as a Maxwell’s Demon that continuously monitors the physical system and appropriately adapts the drift term in order to reduce the physical system’s entropy.

\subsection{Analog Score Network}

We described in Sec.~\ref{sc:entropy_MD} how the Maxwell's Demon can take many physical forms, and the same remarks hold true for the score network. This includes the possibility of using a digital score network in conjunction with an analog s-mode system. However, this can lead to latency issues, whereby the communication between the score network and the s-mode system has some time delay. Hence, we highlight here the possibility of using an analog or hybrid digital-analog architecture for the score network, which could address the latency issue.  

\begin{figure}
\centering
\includegraphics[width=0.9\textwidth]{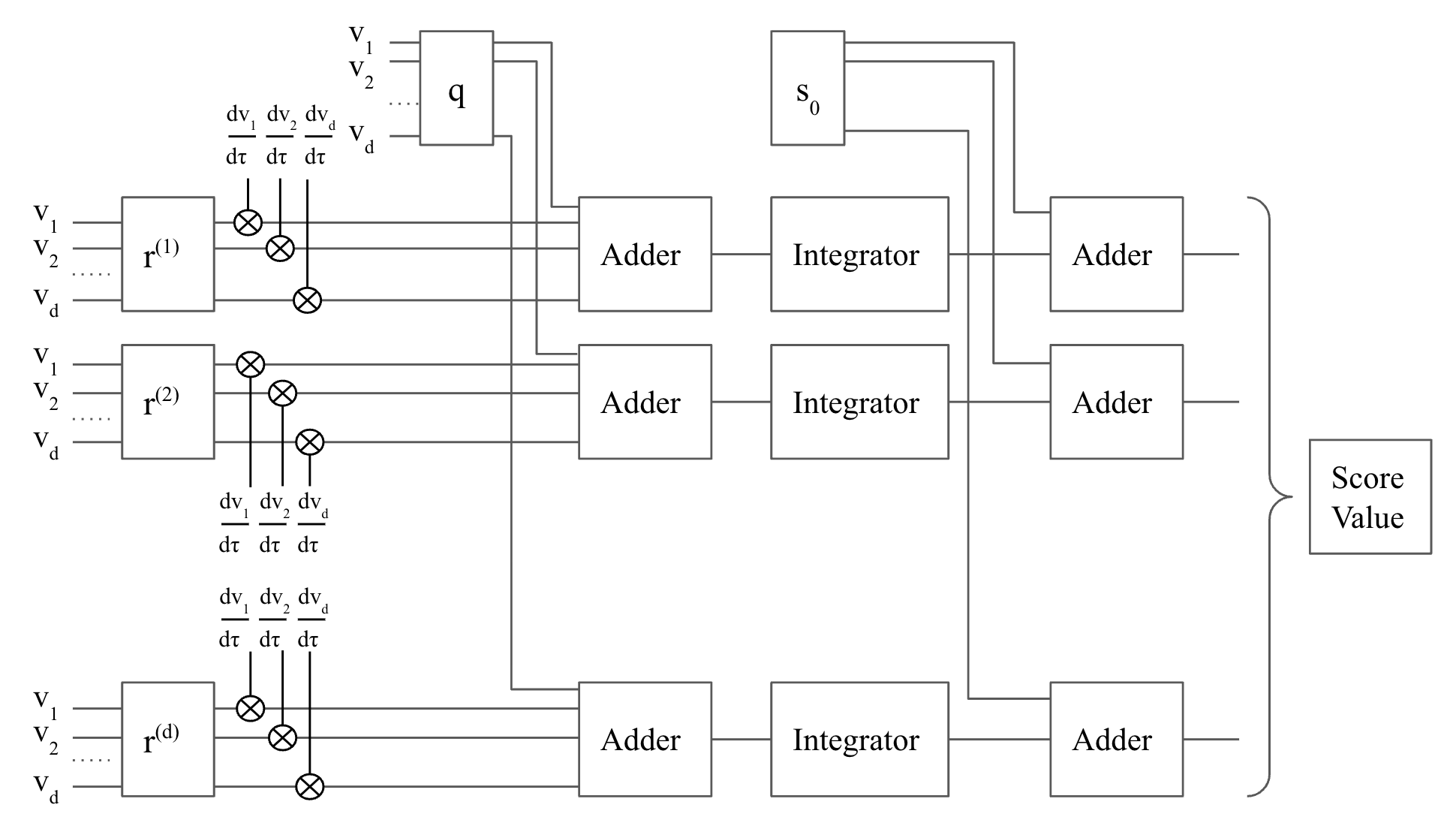}
\caption{\label{fig:scoredevicecircuit}Schematic circuit diagram for an analog score network based on the total derivative approach. This diagram represents the process used to obtain score values during the evolution of the reverse diffusion process.}
\end{figure}

Figure~\ref{fig:scoredevicecircuit} shows an analog circuit for the score network. This circuit is based on the total derivative approach discussed in Sec.~\ref{sc:totalderiv_device}. The subroutines of evaluating $\mathbf{q}$ and $\mathbf{r}$ can be either digital or analog neural networks.

We remark that, instead of the total derivative approach, one could take a force-based approach as described in Sec.~\ref{sc:analogforces}. This would give an alternative means of constructing an analog score network.

\section{Application: Thermodynamic Deep Learning}\label{sc:TDL}

\subsection{Overview}

We introduce Thermodynamic  Deep Learning (TDL) as a broad term for applying Thermodynamic AI Hardware to deep learning. 

Here, we work out details for Bayesian Deep Learning, in particular. Deep learning has been evaluated in high stakes applications, such as self-driving cars and medical diagnosis, where making a wrong decision has major consequences. For these applications, uncertainty quantification (UQ) is crucial, due to the overconfidence of deep learning systems. Bayesian deep learning (BDL) allows for UQ on the predicted output of neural networks. However, it is computationally intractable to perform Bayesian deep learning with both high accuracy and fast speed with current digital hardware~\cite{izmailov2021bayesian}. Typically the mean-field approximation is employed as part of posterior inference although this can lead to inaccuracies~\cite{izmailov2021bayesian}. Thermodynamic AI Hardware could potentially accelerate BDL to the point where large-scale BDL is feasible without having to make such approximations. We further elaborate on this application in the next subsection.

\subsection{Thermodynamic Bayesian Deep Learning}\label{sc:TBDL}

\subsubsection{Background}
Machine learning systems such as neural networks are known to often be overconfident in their predictions. For low-stakes applications, this may not be a major issue. However, some applications, such as self-driving cars and medical diagnosis, are higher stakes in the sense that making a wrong decision can lead to consequences relevant to human life. Overconfidence can be catastrophic for these high-stakes applications.

At a technical level, this overconfidence often arises because neural networks are trained on limited amounts of training data. These training data points live in a vast feature space. Hence it is common for some regions of feature space may not be represented by the training data, i.e., these regions may be far away from the training data. When it comes time to test the trained neural network on testing data, the testing data could be in a region that is far away from training data, and yet the neural network will still attempt to make a prediction in this case. Because the neural network is not familiar with these regions, the neural network is not aware that it should be careful when making predictions for them.

 One strategy for dealing with overconfidence is uncertainty quantification (UQ). UQ aims to quantify the uncertainty of the predictions made by the neural network. UQ is useful for high-stakes applications (e.g., cancer detection in medicine) because it provides guidance for when the user should defer to human judgement over the machine's predictions. UQ is widely recognized as  making machine learning more reliable and trustworthy.

Several different methods exist for UQ in machine learning. A simple example of UQ is adding confidence intervals to the predictions made by the neural network. 

A more sophisticated and rigorous approach to UQ is the Bayesian framework. The Bayesian framework quantifies uncertainty by accounting for prior knowledge (often called the prior distribution) and updates that knowledge due to data or observations (often called the posterior distribution). Bayesian methods aim to quantitatively capture knowledge in the form of probability distributions. 

\subsubsection{Neural Differential Equations}

A continuous-time approach to Bayesian deep learning was recently developed~\cite{xu2022infinitely}, and hence we consider that approach in what follows. 

Neural Ordinary Differential Equations (Neural ODEs)~\cite{chen2018neural} are a continuous depth version of neural networks. The values of the hidden units are denoted $h_t$ and the values of the weights are denoted $w_t$. In general, both of these quantities depend on time, and evolve according to the coupled differential equations:
\[
\frac{d}{dt}
\begin{bmatrix}
h_t\\
w_t
\end{bmatrix}
=
\begin{bmatrix}
f_h(t,h_t,w_t)\\
f_w(t,w_t)
\end{bmatrix}
\]
Hence, a forward pass through the neural network involves integrating this system of differential equations. 

Neural Stochastic Differential Equations (Neural SDEs)~\cite{xu2022infinitely}  are a continuous depth version of Bayesian neural networks. Once again, the system evolves according to a system of coupled differential equations. However, the weights of the neural ODE evolve in a stochastic manner:
\begin{equation}\label{eqn:neuralSDE}
d
\begin{bmatrix}
h_t\\
w_t
\end{bmatrix}
=
\begin{bmatrix}
f_h(t,h_t,w_t)\\
f_w(t,w_t)
\end{bmatrix}
dt +
\begin{bmatrix}
\mathbf{0}\\
g_w(t,w_t)
\end{bmatrix}
dB
\end{equation}
where $dB$ is a Brownian motion term. Equation~\eqref{eqn:neuralSDE} provides the basis for Bayesian deep learning in continuous time.

For the posterior distribution we can specialize Eq.~\eqref{eqn:neuralSDE} to the following form:
\begin{equation}\label{eqn:neuralSDE_posterior}
d
\begin{bmatrix}
h_t\\
w_t
\end{bmatrix}
=
\begin{bmatrix}
f_h(t,h_t,w_t)\\
\text{NN}_{\phi}(t,w_t,\phi)+ w_t
\end{bmatrix}
dt +
\begin{bmatrix}
\mathbf{0}\\
\sigma I_d
\end{bmatrix}
dB
\end{equation}
The posterior distribution needs to be highly expressive. Hence, the SDE must be more complicated than that of the prior distribution. The drift term for the weights is therefore described by a neural network $\text{NN}_{\phi}$ with trainable parameters $\phi$. Note that the trainable parameters in this model include both the parameters $\phi$ appearing in the drift term as well as the initial condition $w_0$ on the weights. The neural network $\text{NN}_{\phi}$ can be referred to as the Posterior Drift Network (PDN), since it determines the drift associated with the posterior distribution.

\subsubsection{Subroutines in Bayesian deep learning hardware}

\begin{figure}
    \centering
\includegraphics[width=0.5\textwidth]{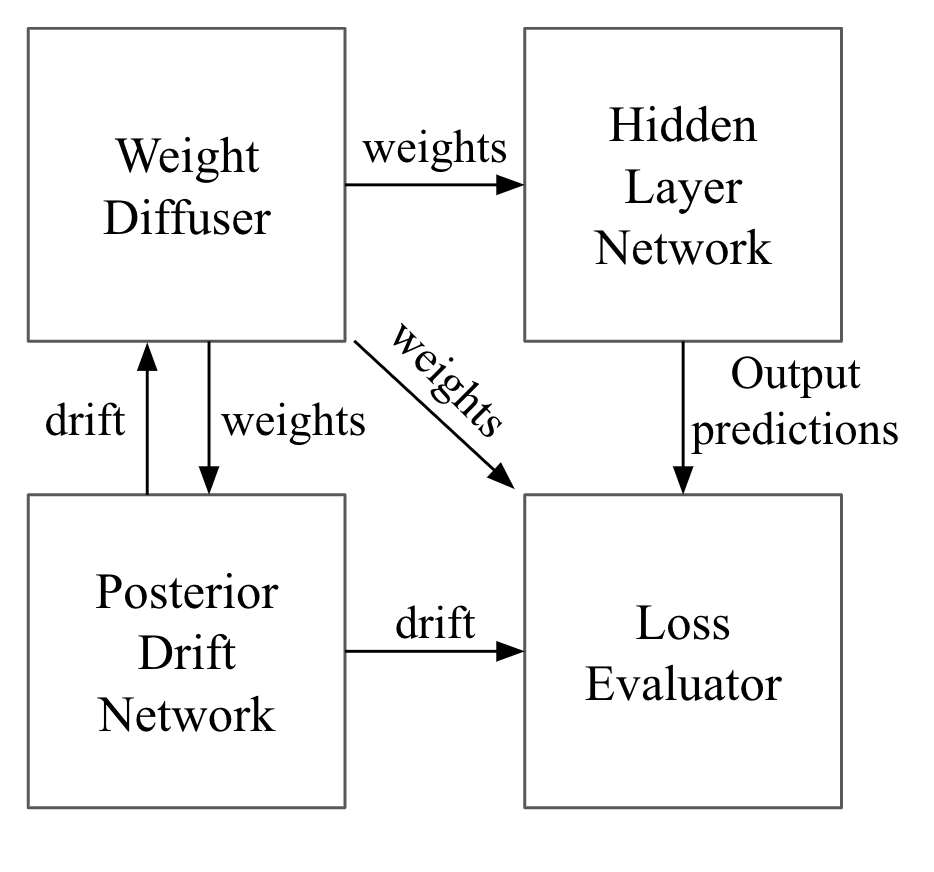}
    \caption{Overview of Thermodynamic Bayesian deep learning, showing how the four subroutines interact and feed signals to each other.}
    \label{fig:BNN_Overview}
\end{figure}

The Thermodynamic AI system for Bayesian deep learning consists of four subroutines. Each of these subroutines can correspond to a physical analog device, or some subset of the subroutines may be stored in and processed on a digital device.  The four subroutines include the following:
\begin{enumerate}
\item Hidden layer network 
\item Weight diffuser
\item Posterior drift network
\item Loss evaluator
\end{enumerate}
Figure~\ref{fig:BNN_Overview} illustrates how these four subroutines interact with each other. The Weight Diffuser (WD), which can represent both the prior distribution and the posterior distribution, feeds weight values to the Hidden Layer Network (HLN). The WD also communicates back-and-forth with the Posterior Drift Network (PDN), in which the WD feeds weight values to the PDN and the PDN feeds drift values to the WD. The Loss Evaluator (LE) takes in signals from all three of the other subroutines - the HLN, the WD, and the PDN - in order to evaluate the loss function.

\subsubsection{Fitting into our Thermodynamic AI framework}

We make the following mapping in order to fit this into our framework:
\begin{align}
    \text{(weight diffuser)} &\leftrightarrow  \text{(s-mode device)}\\
    \text{(posterior drift network)} &\leftrightarrow  \text{(Maxwell's demon device)}
\end{align}
The weight diffuser corresponds to the s-mode device, and the posterior drift network corresponds to the Maxwell's demon device.

The weight diffuser uses s-mode dynamics to sample weight trajectories $w_t$, which are then imported into the HLN. The Maxwell's demon is used to produce complex s-mode dynamics that produce weight trajectories according to a posterior distribution.

\subsubsection{Description of Bayesian Deep Learning Hardware}

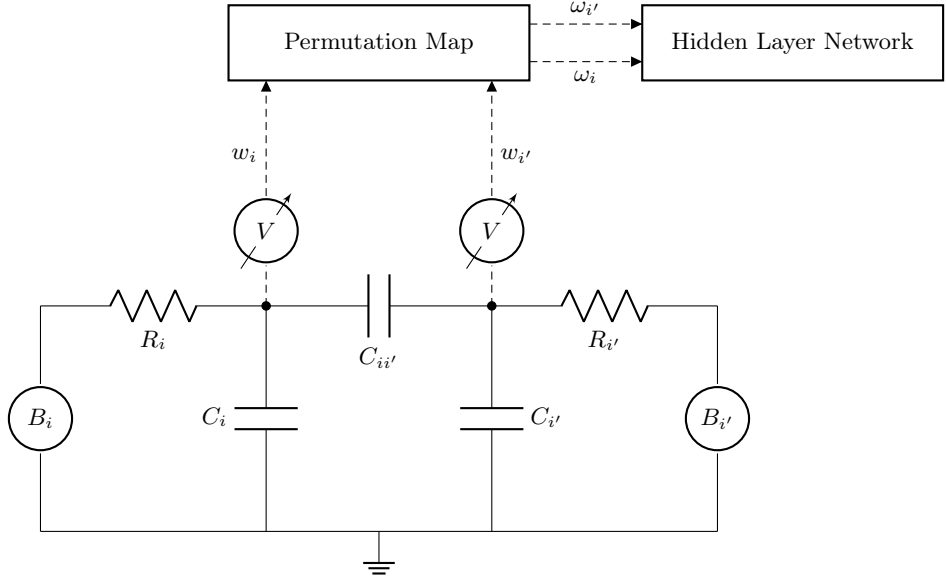
\begin{figure}
    \centering
    \begin{circuitikz}
        \draw (0,1.5) node[rmetershape](b1){} node[]{$B_i$};
        \draw (0,0) -- (b1) to[short] (0,3);
        \draw (0,3) to[R, l_=$R_{i}$, -*] (3,3);
        \draw (3,3) to[C, l_=$C_{i}$] (3,0);
        \draw (3,3) to[C, l_=$C_{ii'}$] (6,3);
        \draw (6,0) to[C, l_=$C_{i'}$] (6,3);
        \draw (6,3) to[R, l_=$R_{i'}$, *-] (9,3);
        \draw (9,1.5) node[rmetershape](b2){} node[]{$B_{i'}$};
        \draw (9,0) -- (b2) to[short] (9,3);
        \draw (0,0) to[short] (4.5,0) node[ground]{} -- (9,0);
        \draw (3,4) node[rmeterwashape](v1){} node[]{$V$};
        \draw (6,4) node[rmeterwashape](v2){} node[]{$V$};
        \draw [densely dashed] (3,3) -- (v1) (6,3) -- (v2);
        \draw [thick] (2.5,6) rectangle (6.5,7) node[pos=0.5]{Permutation Map};
        \draw [densely dashed] (v1) -- (3,6) node[inputarrow, rotate=90]{};
        \draw [densely dashed] (v2) -- (6,6) node[inputarrow, rotate=90]{};
        \draw (3,5) node[left]{$w_i$} (6,5) node[right]{$w_{i'}$};
        \draw [densely dashed] (6.5,6.25) -- (8,6.25) node[inputarrow]{};
        \draw [densely dashed] (6.5,6.75) -- (8,6.75) node[inputarrow]{};
        \draw (7.25,6.25) node[below]{$\omega_i$} (7.25,6.75) node[above]{$\omega_{i'}$};
        \draw [thick] (8,6) rectangle (12,7) node[pos=0.5]{Hidden Layer Network};
    \end{circuitikz}
    \caption{\label{fig:OutputtingWeights}Outputting weights from the weight diffuser device to the HLN device. Shown here is the simplified case for two unit cells of the weight diffuser, although the concept applies to an arbitrary number of unit cells. A permutation map (which physically correspond to a wire routing scheme) is shown to provide flexibility for how the weights $\mathbf{w}$ outputted by the WD get incorporated as weights $\mathbf{\omega}$ in the HLN.} 
\end{figure}

Figure~\ref{fig:OutputtingWeights} illustrates a possible hardware implementation of the weight diffuser. As shown, the outputs of the weight diffuser are supplied to the Hidden Layer Network device. 

The weight diffuser can be an analog circuit, with variables $v(t)$, equal the voltage values across a set of capacitors located in a series of unit cells. These voltage values diffuse according to a stochastic process inside the circuit with analog noise sources $B_i$, and are output to a hidden layer network as time-continuous weight trajectories. The same diffusing cell can generate prior and posterior samples, depending on whether the posterior drift network applies a demon voltage vector to the circuit.  

The hidden layer network may also be an analog circuit, whose dynamics can be modeled as a neural ODE dependent on the sampled weight trajectory outputted by the weight diffuser.

\section{Application: Thermodynamic Monte Carlo}\label{sc:montecarlo}

\subsection{Background}
Monte Carlo algorithms~\cite{metropolis1987beginning} have become widely used, finding application in finance, physics, chemistry, and more recently, machine learning~\cite{kroese2014monte}.  Monte Carlo algorithms provide a simple procedure for approximating integrals involving probability distributions.  Suppose we have a probability distribution $\pi(\mathbf{x})$ with $\mathbf{x}\in \mathcal{X}$ the sample space, and suppose we want to find the expectation value of some function $f(\mathbf{x})$ with respect to $\pi$.  The Monte Carlo method consists in approximating the integral with a simple sample average, 
\begin{equation}
    \int f(\mathbf{x}) \pi(\mathbf{x}) d\mathbf{x} \approx \frac{1}{M}\sum_{i=1}^M f(\mathbf{x}_i),
\end{equation}
where the samples $\mathbf{x}_i$ are distributed according to $\pi$.  The computational bottleneck has been transformed from integration to sampling.  The best methods for sampling from $\pi$ will depend on the form of $\pi$.

Markov Chain Monte Carlo (MCMC) is one popular strategy for constructing samplers \cite{metropolis1953equation}, and can be applied whether the state space $\mathcal{X}$ is discrete or continuous.  This strategy involves setting up a chain of dependent samples, such that over a long enough time, the set of samples becomes distributed according to $\pi(\mathbf{x})$.  In other words, it operates by constructing a Markov chain that has the target distribution $\pi$ as its stationary distribution.  An MCMC algorithm is defined by making a choice of conditional distribution $g(\mathbf{x}'|\mathbf{x})$ and a choice of initial state $\mathbf{x}_0$, then repeating the following for each iteration $i$ \cite{voss2013introduction}:
\begin{enumerate}
    \item Initialize at position $\mathbf{x}_i$
    \item Draw a sample from $g$ as $\mathbf{x}' \sim g(\mathbf{x}'| \mathbf{x}_i)$
    \item Compute the probability $\Pi_a$ of accepting $\mathbf{x}'$ as the next state in the chain.  The probability is given by 
    \begin{equation}
    \Pi_a(\mathbf{x}', \mathbf{x}_i) = \min \left( 1, \frac{\pi(\mathbf{x}')g(\mathbf{x}_i| \mathbf{x}')}{\pi(\mathbf{x}_i)g(\mathbf{x}'| \mathbf{x}_i)}\right)
    \end{equation}
    \item Draw a random number $r$ uniformly from the unit interval $[0,1]$. If $r<\Pi_a$, accept the move by setting $\mathbf{x}_{i+1} = \mathbf{x}'$.  Otherwise, stay at the current position by setting $\mathbf{x}_{i+1} = \mathbf{x}_i$.
\end{enumerate}
Note that one should choose the conditional distribution $g$ such that it is easier to sample from than the target distribution $\pi$.

In the following sections we will present how the framework of thermodynamic AI systems naturally encompasses Monte Carlo algorithms, focusing on two key algorithms: Langevin Monte Carlo (LMC) and Hamiltonian Monte Carlo (HMC)~\cite{betancourt2017conceptual}.

\subsubsection{Hamiltonian Monte Carlo}
\begin{figure}
    \centering
    \includegraphics[scale=0.35]{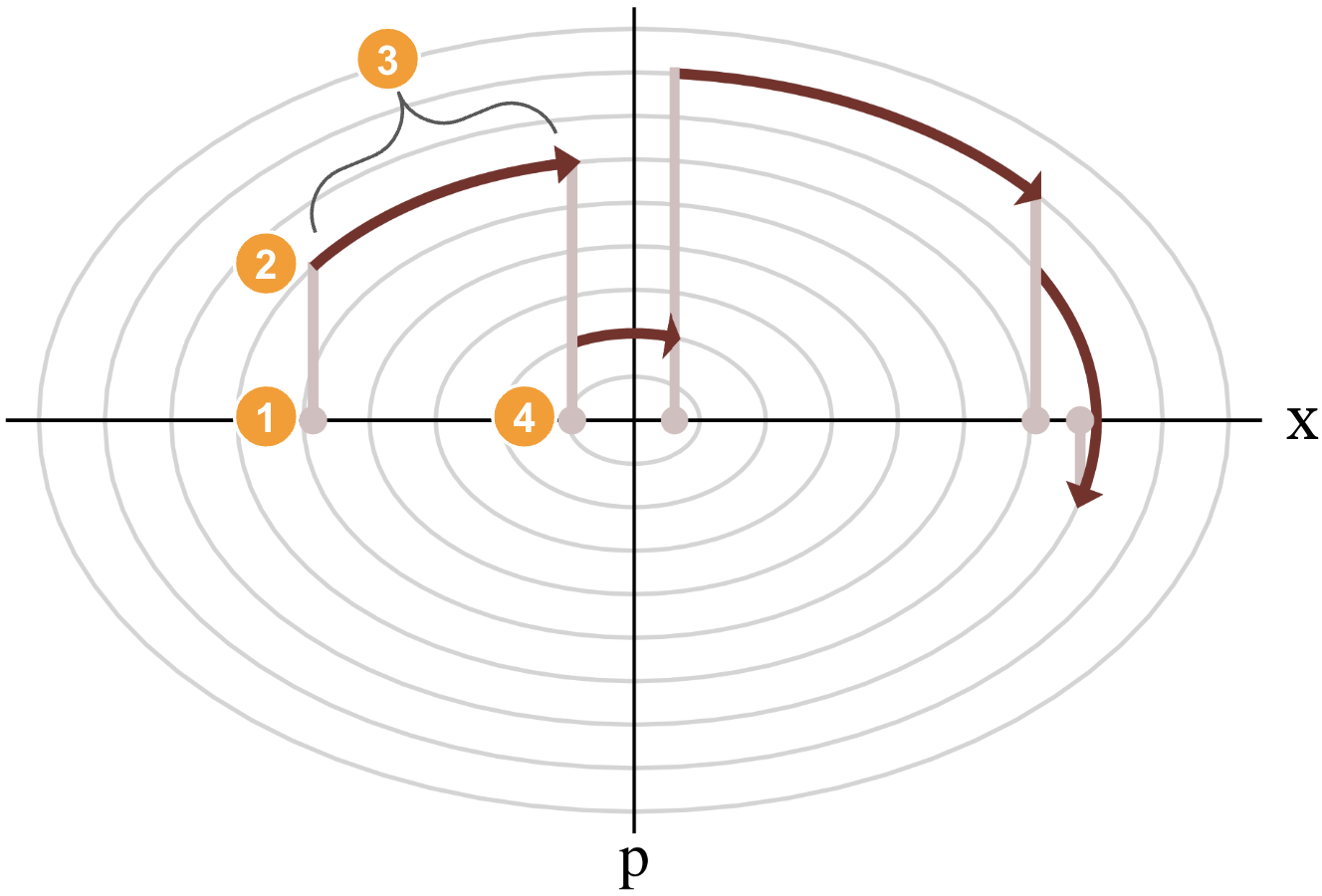}
    \caption{Phase space picture of the HMC algorithm, for a 1D parabolic potential. Step 1: Initialize at the most recent chain position $x_i$. Step 2: Randomly sample momentum $p$ from a known probability distribution. Step 3: Integrate Hamilton's equations. Step 4: accept or reject the move from $x_i$ to $x'$. Figure adapted from~\cite{betancourt2017conceptual}.}
    \label{fig:hmc_pic}
\end{figure}
HMC has gradually become one of the most widely used MCMC algorithms for statistical analysis and learning  thanks to its computational efficiency and sample quality \cite{neal2011mcmc}.  The markov chain in HMC is constructed by proposing new samples using a combination of gradient information and Hamiltonian dynamics.
The key idea behind HMC is to introduce fictitious momentum variables $\mathbf{p}$, to which $\mathbf{x}$ is coupled according to Hamilton's equations:
\begin{align}
    \frac{\partial \mathbf{x}}{\partial t} =&\frac{\partial}{\partial \mathbf{p}} H(\mathbf{x},\mathbf{p}) \\
    \frac{\partial \mathbf{p}}{\partial t}=& -\frac{\partial}{\partial \mathbf{x}} H( \mathbf{x},\mathbf{p})
\end{align} with the Hamiltonian $H$ defined as $H( \mathbf{x},\mathbf{p}) = -\log\pi( \mathbf{x}) + \mathbf{p}^TM\mathbf{p}/2$, with a potential term $U(\mathbf{x})=-~\log\pi(\mathbf{x})$ corresponding to the landscape of the target probability distribution in log space, and $M$ a mass matrix. At each iteration of the HMC algorithm, a new sample is proposed by integrating the equations of motion in phase space over a fixed time interval $\tau$. As such, HMC is referred to as a gradient-augmented MCMC method, where information on the gradient of the log-probability is integrated in the chain. The proposed sample is then accepted or rejected using the Metropolis-Hastings acceptance criterion, which in the case of Hamiltonian dynamics reduces to 
\begin{equation}
    \Pi_a(\mathbf{x}', \mathbf{x}) = \frac{\pi(\mathbf{x}')}{\pi(\mathbf{x})}
\end{equation} since the dynamics are reversible which gives $g(\mathbf{x}',\mathbf{x}) = g(\mathbf{x}, \mathbf{x}')$.
The gradient information in HMC is used to define the direction of the proposed updates, allowing the Markov chain to efficiently explore regions of high probability mass. Therefore,  regions of low probability mass may be avoided, thus allowing the Markov chain to escape from local modes and explore the target distribution more effectively unlike random walk MCMC methods. One may already see the connection with force-based MD, introduced in section IX, that is made more explicitly in the next subsection.

A more elaborate version of HMC has been developed and is also widely used in statistics, coined the No U-Turn sampler (NUTS). NUTS has the advantage of automatically tuning the step size and the trajectory during the sampling process, making it easier to use in practice~\cite{hoffman2014no}.
\subsubsection{Stochastic Gradient Hamiltonian Monte Carlo}
Stochastic Gradient Hamiltonian Monte Carlo (SGHMC)~\cite{chen2014stochastic} is an extension of HMC, proposed to use HMC efficiently on large problem sizes where computing exactly the gradient of the log probability $\nabla_{\mathbf{x}} \log \pi(\mathbf{x}) = \nabla U(\mathbf{x})$ (which is necessary to compute the dynamics) cannot be performed. Indeed, this gradient can be expressed as 
\begin{equation}
    \nabla U(\mathbf{x}) = - \sum_{\mathbf{x}_i \in \mathcal{D}} \nabla \log \pi(\mathbf{x}_i|\mathbf{x}) - \nabla\log p(\mathbf{x}).
\end{equation} for points $\mathbf{x}_i \in \mathcal{D}$, with $\mathcal{D}$ the set of observations. For large problem sizes, this quickly becomes intractable. To overcome this, the gradient may be approximated by uniformly sampling points $\mathbf{x}_i \in \tilde{\mathcal{D}}, \tilde{\mathcal{D}} \subset \mathcal{D}$:
\begin{equation}
    \nabla \tilde{U}(\mathbf{x}) = -\frac{|\mathcal{D}|}{|\tilde{\mathcal{D}}|} \sum_{\mathbf{x}_i \in \tilde{\mathcal{D}}} \nabla \log \pi(\mathbf{x}_i|\mathbf{x}) - \nabla \log p(\mathbf{x}).
\end{equation} Assuming the $x_i$ are independent, the central limit theorem leads to:
\begin{equation}
    \nabla \tilde{U}(\mathbf{x}) \approx \nabla U(\mathbf{x}) + \mathcal{N}(0,V(\theta))
\end{equation} with $V$ the covariance of Gaussian noise with zero mean coming from the stochastic gradient approximation. In ref.~\cite{chen2014stochastic}, it has been shown that one can add a friction term to counterbalance the effect of the noise coming from the stochastic gradient, thus obtaining the  dynamical equations for SGHMC with friction:
\begin{align}\label{eq:sghmc_111}
    \mathrm{d}\mathbf{x} &= M^{-1} \mathbf{p}\dt\\
    \mathrm{d}\mathbf{p} &= -  [\nabla U(\mathbf{x}) + BM^{-1}\mathbf{p}] \dt + \sqrt{2}B\mathrm{d}\mathbf{w} \label{eq:sghmc}
\end{align}where $M$ is a mass matrix, $B= V(\theta)/2$ is the diffusion matrix, $\nabla U(\mathbf{x}) = \mathbf{f}(\mathbf{x})$ is the force and $\mathbf{w}$ is the Wiener process. The stationary distribution for $\mathbf{x}$ obtained at long times corresponds to the target probability distribution $\pi(\mathbf{x}) = \exp\{-U(\mathbf{x})\}$. In fact, looking at Eq.~\eqref{eq:sghmc}, one can see the close connection with Eqs.~\eqref{eqn:VPVEunified2} and~\eqref{eqn:neuralSDE_posterior}.

\subsection{Connection to Langevin Monte Carlo}

There is a close connection between SG-HMC and Langevin Monte Carlo (LMC), in particular with a variation of LMC known as stochastic gradient Langevin dynamics (SGLD).  SGLD is a procedure for Bayesian posterior sampling of the parameters of a machine learning model.  As in  \cite{welling2011bayesian}, let $\theta$ be the parameters of the model, let $p(\theta)$ be a prior distribution on the parameters, and let $p(x|\theta)$ be the probability of data point $x$ given that our model is parameterized by $\theta$.  Similarly to SGHMC, we can imagine introducing the $N$ data points randomly in small batches of size $n$.  Then the SGLD dynamics are specified by the update equation
\begin{equation}
    \Delta \theta_t = \frac{\epsilon_t}{2}\left(\nabla \log p(\theta_t) + \frac{N}{n} \nabla \log p(x_{ti}|\theta_t)\right) + \mathcal{N}(0, \epsilon_t),
\end{equation}
where $\epsilon_t$ is a time-dependent step size. The noise term prevents the the parameters from freezing at a particular value, instead being spread according to the posterior distribution.  From this perspective we can see thermodynamic fluctuations as a resource for posterior inference.

We can further cast this equation in terms of the force framework introduced in section \ref{sc:entropy_MD}.  Since we can treat the logarithm of a distribution as an energy, we can write $U(t, \theta) = \log p(\theta) + \frac{N}{n}\log p(x_{ti}|\theta_t)$, so that $U(t, \theta)$ is a time dependent energy function. Then we have
\begin{equation}
    \mathrm{d} \theta = \frac{\epsilon_t}{2}\nabla U(t, \theta)\mathrm{d}t + \sqrt{\epsilon_t}dW.
\end{equation}
Thus the data becomes part of a time dependent diffusion vector.

\subsection{Fitting into our Thermodynamic AI Framework}

Figure~\ref{fig:tmc_device} shows how these algorithms can fit into our framework. Referring to the terminology introduced in section~\ref{sc:analogforces}, we propose the following mapping to the s-unit formalism:\begin{align}
    \text{(momentum device)} &\leftrightarrow  \text{(s-mode device, with a noise source and an injected value for the force)}\\
    \text{(position device)} &\leftrightarrow  \text{(latent variable stored in the MD's memory)}
\end{align}
This mapping refers to how the differential equations in, for example, Eq.~\eqref{eq:sghmc_111} and Eq.~\eqref{eq:sghmc} are mapped to the devices in the Thermodynamic AI system.
Assuming the noise is uncorrelated, one can set $B$ in Eq.~\eqref{eq:sghmc} to a scalar value. $M$ will therefore be given by the connectivity of the momentum device.

\subsection{Description of Monte Carlo Hardware}

The coupled differential equations described above are amenable to being implemented on Thermodynamic AI hardware. 
Implementing the SGHMC algorithm on digital hardware requires one to compute the derivatives of position and momentum, which involve diagonalizing matrices, hence having a computation cost in $O(n^3)$ in the general case, with $n$ the number of data points.

By implementing SGHMC in Thermodynamic AI hardware, this scaling may be eased. This can help in alleviating the bottleneck that is sampling for many applications. We coin Thermodynamic Monte Carlo (TMC) for the implementation of Monte Carlo algorithms in Thermodynamic AI hardware.

For the hardware implementation of SGMHC, one can consider two devices: one whose state variable is $\mathbf{x}$, and one whose state variable is $\mathbf{p}$. Ultimately, one is interested in obtaining values of $\mathbf{x}$, whose evolution is dictated by the dynamics of $\mathbf{p}$. To unify the SGHMC and SGLD approaches, we propose a single hardware paradigm. 

\begin{figure}[t]
    \centering
    \includegraphics[scale=0.45]{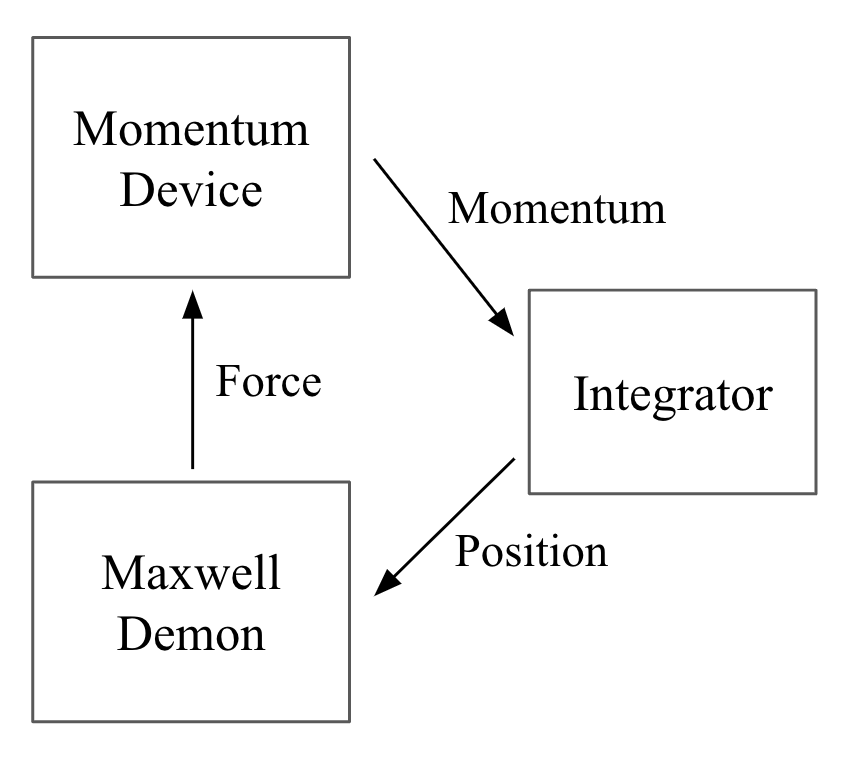}
    \caption{Schematic diagram for a Thermodynamic Monte Carlo device. The momentum device corresponds to an s-mode device, with s-modes constructed as described in Sec.~\ref{sc:phys_smode}. The Maxwell's demon is constructed via the force-based approach discussed in Sec.~\ref{sc:analogforces}. }
    \label{fig:tmc_device}
\end{figure}

This platform is depicted in Fig.~\ref{fig:tmc_device}. The force is calculated by the MD, which is then fed into the momentum device. The momentum vector is fed in real time to the integrator, the result of which is the position vector which is fed back into the Maxwell's demon for storage.

In the case of SGHMC, a noisy estimate of the force is calculated, which is then fed into the momentum device, which has to have no Brownian noise so that the correct SDE is implemented. The friction term may be implemented by dissipative elements, such as resistors.

In the case of SGLD, the force is calculated exactly, and then fed into the momentum device that will have both friction (that is high in this case) and Gaussian noise.

\section{Application: Thermodynamic Annealing}\label{sc:annealing}

\subsection{Background}

Many important problems can be phrased as optimization problems. Posing a problem as an optimization problem means defining a loss function on the space of potential solutions, such that the better the answer, the lower the corresponding value of the loss function.
While many methods exist for solving optimization problems, a key difficulty that must be overcome is the existence of local minima in the loss function.  These local minima show up whether the solution space is discrete or continuous.  We outline a naive way to approach each domain, and how local minima thwart successfully finding the global optimum:
\begin{itemize}
\item For continuous problems, we might apply gradient descent \cite{cauchy1847methode}.  Starting from a random location in solution space, the algorithm repeatedly computes the local gradient and uses it to travel down hill.  If the algorithm reaches a local minimum, the gradient becomes zero, so the algorithm ceases to move.
\item For discrete problems, we might apply local search \cite{dunham1964design}.  Starting from a random configuration, we compute the value of all neighboring configurations; if any neighbor has a lower loss function value, the algorithm moves to that configuration.  In a local minimum, no neighbor has a lower value of the loss, halting progress.
\end{itemize}
In both of these cases, we need something more to prevent becoming trapped in a local minimum.

\begin{figure}
    \centering
\includegraphics[width=0.75\textwidth]{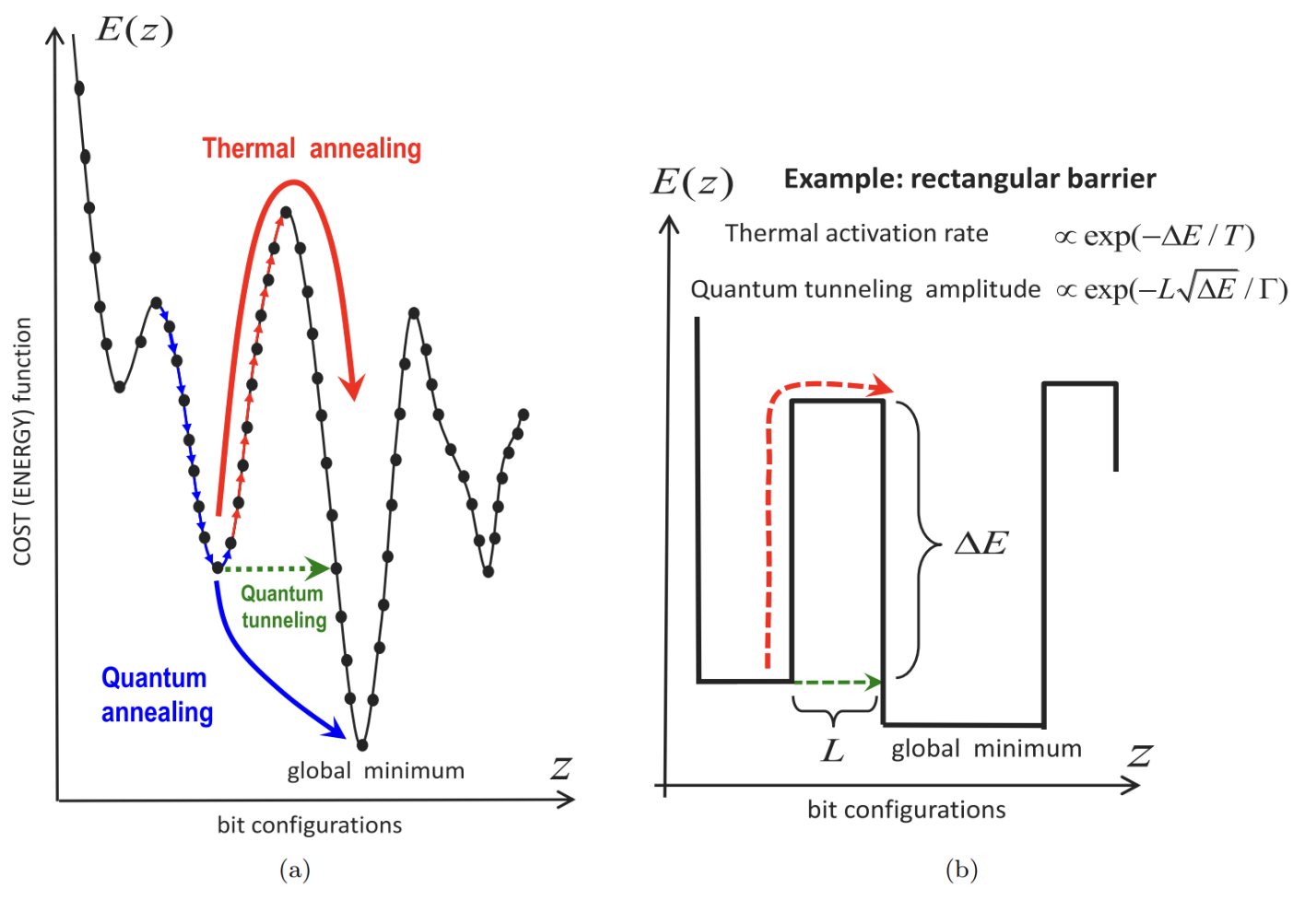}
    \caption{Illustration of thermal and quantum fluctuations enabling escape from local minima.  (a) Thermal fluctuations provide temporary energy boosts that enable climbing over barriers.  Quantum fluctuations make it possible to tunnel through barriers. (b) For a simple rectangular barrier, we can write down the transition probabilities provided by both thermal and quantum fluctuations.  We see that they scale differently: thermal fluctuations do not depend on the width of the barrier, while quantum tunneling does. Figure from Ref.~\cite{smelyanskiy2012near}.}
    \label{fig:LocalMinima}
\end{figure}

The key missing ingredient is the ability to temporarily move to a solution or configuration that has a \textbf{larger} value of the loss.  After such a temporary up-hill traversal, the algorithm has a chance to move into a neighboring, deeper minimum.  This is where we can leverage thermal fluctuations as a resource for computation.  If one perturbs the gradient direction (in continuous problems) or the loss values of the neighbors (for discrete problems), then the transition step no longer gets stuck in local minima.  An illustration of this thermal escape from a minimum is illustrated in Figure~\ref{fig:LocalMinima}.

On digital systems, one algorithm which takes advantage of thermal fluctuations is Simulated Annealing, initially proposed in 1983 \cite{kirkpatrick1983optimization}.  Annealing is a process in which a metal is slowly cooled from a high temperature to increase its strength.  Simulated Annealing makes an analogy between such annealing of a metal and optimization: the bonds in the metal are analogous to the loss function, while the strengthening of the metal is analogous to finding a better optimum.

In a physical system, all these perturbations can be provided by thermal fluctuations.  In the next section we formalize the connection between the stochastic dynamics of a physical system and the Simulated Annealing algorithm.

\subsection{SDE approach to simulated annealing}

Reference~\cite{batou2019approximate} provides a mathematical framework for simulated annealing based on SDEs. Let us discuss that framework now.

Suppose that we have an optimization problem, where the loss function $\mathcal{L}(\mathbf{x})$ of interest is a continuous function of an N-dimensional state variable $\mathbf{x}$. Here, $\mathbf{x}$ is the variable that one is optimizing over to solve the optimization problem in the context of simulated annealing. In this setting, one can propose a coupled system of equations for the state variable $\mathbf{x}(t)$ and an auxiliary variable $\mathbf{p}(t)$: 
\begin{align} \label{annealing_SDE1}
    d\mathbf{x}(t) &= \mathbf{p}(t)dt \\
    \label{annealing_SDE2}
d\mathbf{p}(t) &= - \nabla \mathcal{L}(\mathbf{x})dt - \frac{1}{2}D \, \mathbf{p}(t)dt + S \, d\mathbf{W} 
\end{align}
Here, $\mathbf{W}$ is an N-dimensional Brownian motion, $S$ an $N \times N$ dimensional lower-triangular matrix, and $D = SS^{\intercal}$. We call the first differential equation the optimization ODE, while we call Eq.~\eqref{annealing_SDE2} the auxiliary SDE.  

The dynamics on the state variable $\mathbf{x}(t)$ are effectively stochastic, as it is coupled to the auxiliary variable evolving via the auxiliary SDE. 

In the long-time limit, the state variable $\mathbf{x}(t)$ is distributed according to a Boltzmann probability distribution constructed from the loss function $\mathcal{L}$:
\begin{equation}
    \mathbf{x}(t \to \infty) \sim \mathrm{exp}(-\mathcal{L}(\mathbf{x}))
\end{equation}
As such, the long-run samples will most often be concentrated around the extrema of the loss function $\mathcal{L}$, allowing one to identify the minima or maxima. As the state of $\mathbf{x}(t)$ is probabilistic, the entire extrema landscape of $\mathcal{L}$ can be explored.

\subsection{Fitting into our Thermodynamic AI framework}

Equations~\eqref{annealing_SDE1} and~\eqref{annealing_SDE2} fit into our framework for Thermodynamic AI hardware. Specifically, we have the following mapping to our hardware:
\begin{align}
    \text{(auxiliary SDE)} &\leftrightarrow  \text{(s-mode device)}\\
    \text{(optimization ODE)} &\leftrightarrow  \text{(latent variable evolution in Maxwell's demon device)}
\end{align}
The idea is that the auxiliary SDE describing the evolution of $\mathbf{p}$ can be performed on the s-mode device. Here, $S$ would correspond to the coefficient $C(t)$ in our hardware, and $-(1/2)D$ would correspond to the coefficient $A(t)$ in our hardware.

In addition, $- \nabla \mathcal{L}(\mathbf{x})$ would correspond to the demon vector $\mathbf{d}$ in our hardware. The optimization ODE then maps onto the evolution of the latent variable in the Maxwell's demon device. Note that this employs the framework discussed in Sec.~\ref{sc:analogforces} involving a forced-based Maxwell's demon.
Also, note that the mass matrix that appears in our framework is set to be the identity for this application: $M = I$.

\section{Application: Time Series Forecasting}\label{sc:timeseries}

\subsection{Background}

As a final application, we consider analysis of time-series data. Time-series data provide an important application relevant to financial analysis, market prediction, epidemiology, and medical data analysis. In many case one has data at particular time points that may be at irregular time intervals, and one wishes to have a model that makes a predictions at all times and hence one that interpolates between the datapoints. In addition, one may want to a model that extrapolates beyond the data, e.g., to make predictions about the future where no data is available.

Discrete neural networks, such as recurrent neural networks, have been used in the past for interpolating and extrapolating time-series data. However, latent ordinary differential equations (latent ODEs)~\cite{chen2018neural} have been shown to outperform recurrent neural networks at this task. One can view a latent ODE as a parameterized ODE, where the parameters are trained in order to fit the time-series data (according to some loss function). More recently, latent SDEs have been explored for fitting and extrapolating time-series data~\cite{li2020scalable}.

\subsection{Fitting into our Thermodynamic AI framework}

In what follows, we discuss using Thermodynamic AI hardware as either a latent ODE or latent SDE, in order to interpolate and extrapolate a time-series dataset.

For concreteness, consider the case of a latent SDE. In this case, the idea is that the SDE should have trainable parameters that allow it to be fit to the data. This fits well with our Thermodynamic AI hardware, since one can use an s-mode device combined with a (parameterized) Maxwell's demon device to generate a parameterized SDE. For example, the overall dynamics associated with this parameterized SDE could be given by Eq.~\eqref{eqn_fullyprogrammed_SDE}, repeated here for convenience:
\begin{equation}\label{eqn_latentSDE}
    d\mathbf{v}(t) = (A(t) \mathbf{v}(t)+\mathbf{b}(t)+ D(t)\mathbf{d}_{\theta}(t, \mathbf{v}(t)))dt + C(t) d\mathbf{w}\,.
\end{equation}

\begin{figure}
\centering
\includegraphics[width=0.8\textwidth]{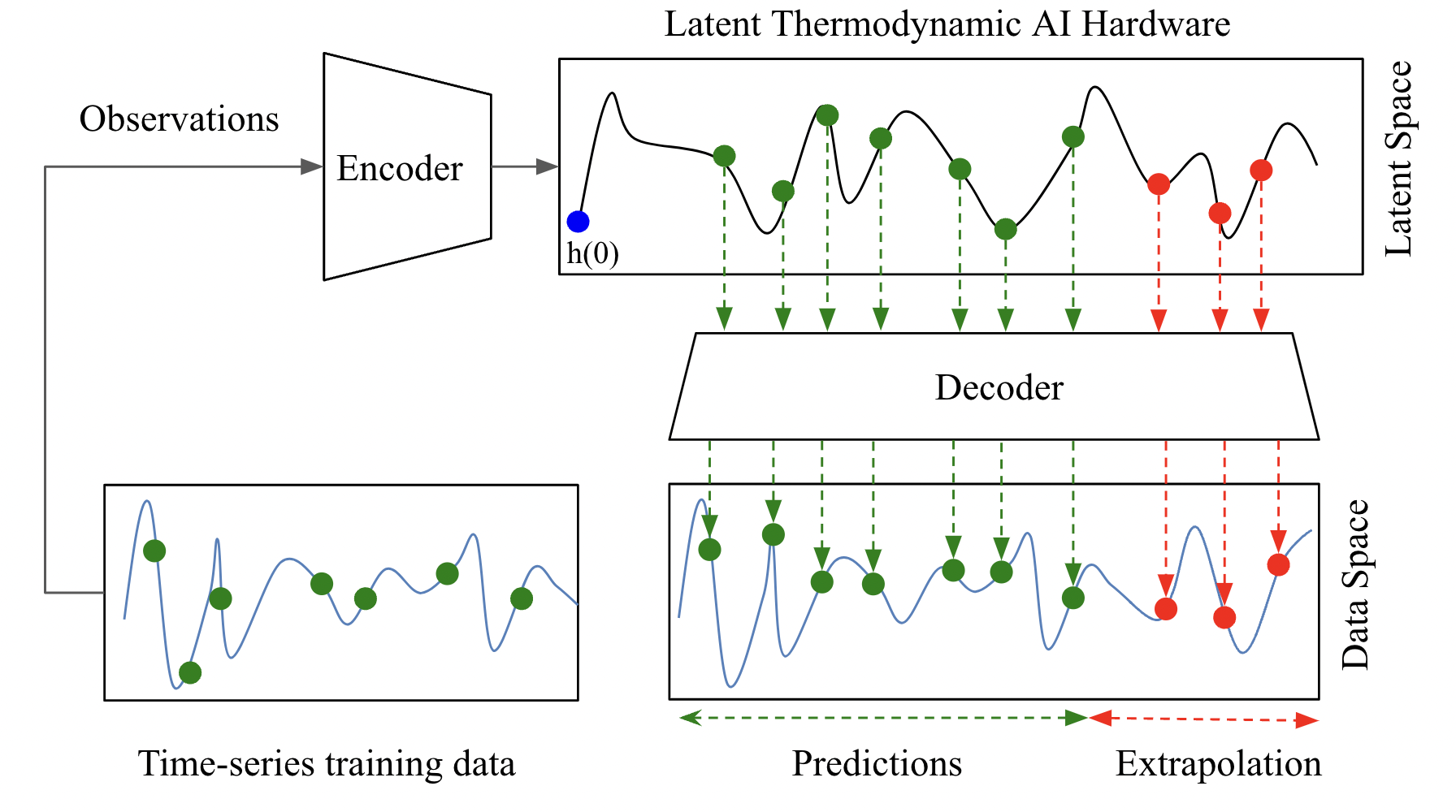}
\caption{\label{fig:LatentThermo} Illustration of  latent Thermodynamic AI hardware for fitting and extrapolating time-series data.}
\end{figure}

Figure~\ref{fig:LatentThermo} provides a schematic diagram for a potential approach. The overall model in Fig.~\ref{fig:LatentThermo} has three subroutines:
\begin{enumerate}
    \item Encoder
    \item Latent Thermodynamic AI hardware
    \item Decoder
\end{enumerate}

The training data are provided as observations from some time series. These time-series observations are fed into an encoder. The encoder has free parameters that can be trained. For example, the encoder could be a recurrent neural network. The output of the encoder can be the initial vector $\mathbf{h}(0)$ of the hidden layer values, or the output could be a probability distribution from which $\mathbf{h}(0)$ is sampled. If the encoder is stored on a digital device, its output can converted to an analog signal.

The Thermodynamic AI hardware acts as the latent space for the latent ODE or latent SDE. This latent space is initialized to $\mathbf{h}(0)$ by the encoder. Then then hidden layer values evolve over time according to ODE or SDE that describes the system, such as the SDE in Eq.~\eqref{eqn_latentSDE}.

The hidden layer values $h(t_k)$ can be read off at a set $\{t_k\}$ of various times, e.g., by measuring the state variables of the s-mode system in the Thermodynamic AI hardware. This set $\{h(t_k)\}$ of values can be fed to a decoder. The decoder can have free parameters that will be trained. The outputs of the decoder correspond to predictions that the latent ODE model makes for the true time series. These predictions can go beyond the time interval associated with the observations, in which case the predictions correspond to extrapolated values.

A training process occurs where the parameters of the encoder, of the decoder, and of the Maxwell's demon in the Thermodynamic AI hardware optimized in order to minimize or maximize a loss function. This essentially corresponds to fitting the time-series data. Gradient based approaches such as the adjoint sensitivity method can be employed here.

\section{Thermodynamic Speedups}

Quantum computers are known to give speedups on certain problems like quantum simulation - these are so-called quantum speedups. By analogy, we coin the term thermodynamic speedups for the case when Thermodynamic AI hardware accelerates an algorithm relative to standard, digital hardware. There are several avenues by which one could obtain a thermodynamic speedup.

For example, Feynman~\cite{feynman1982simulating} noted that joint probability distributions are typically hard to marginalize on digital devices, but when such distributions are prepared on physical devices with spatially separated components, then one gets the marginal distributions for free - without any computation. While Feynman was discussing quantum computers, this argument also applies to Thermodynamic AI hardware. Hence, marginalizing joint probability distributions occurs naturally on Thermodynamic AI hardware.

A second source of speedup really gets to the heart of Thermodynamic AI hardware by viewing stochasticity as a resource. Many AI algorithms involve probabilistic variables, yet must be implemented with zero-entropy states when solved on digital computers. In general, noisy physical systems provide access to states with non-zero entropy, so probabilistic aspects are injected naturally (instead of with computational effort) whenever these states are the object of computation. In other words, producing stochasticity on digital hardware is a subroutine that takes computational effort, whereas no such subroutine is necessary for algorithms run on Thermodynamic AI hardware.

A third source of speedup arises from the Maxwell demon. Oftentimes, AI algorithms formulate the mathematics of the Maxwell's demon in the language of physics, involving the gradient of the potential energy function and phase-space dynamics. These subroutines are done by brute force on digital hardware. However, by making the Maxwell's demon physical, the phase space dynamics would naturally occur (as opposed to requiring digital integration). Perhaps more importantly, the gradient of the potential energy is a physical force, as illustrated in Fig.~\ref{fig:Potential}. Hence, this raises the possibility of simply measuring this force instead of computing it. The potential $U_{\theta}(\mathbf{x},t)$ may have a complex landscape, yet the physics of the system would ensure the gradient of this potential can be accessed instantaneously.

A fourth source of speedup comes during the training process. Many loss functions are formulated in terms of subroutines that analog hardware can speedup, such as computing vector norms or computing time integrals. For example, this is the case for score-matching loss functions in diffusion models~\cite{song2020score}. Accelerating the estimation of the loss function or its gradient during training thus offers another avenue for speedup.

As a final source of speedup, one can delve into the SDE integration subroutine. In Thermodynamic AI hardware, SDE integration occurs naturally in the s-mode (or s-bit) device. No matrix-vector multiplications are needed, no discretization of time is needed, and no numerical instabilities occur when time-evolving the s-mode device. This has the following implications:
\begin{enumerate}
    \item Digitally calculating the terms on the right-hand side of the SDE (e.g., in Eq.~\eqref{eqn_fullyprogrammed_SDE}) may be computationally demanding. It involves many matrix multiplications and inversions, in some cases seen throughout the article. By solving the SDEs on physics-based hardware, one naturally avoids this problem.
    \item SDEs are in general numerically unstable for large dimensions, requiring sophisticated numerical integrators and fine-tuned time-stepping schedules. With a physics-based hardware system, these difficulties disappear, as there is no time step to be chosen. 
    \item Unlike digital approaches, the total physical time to solve the various SDEs can be tuned and depends on the operating decay rate of the system, proportional to $1/(RC)$ for a single s-mode based on an RC circuit. This means the physical time of integration can be tuned, and when possible can be made faster, which is an advantage of physics-based computation in general. 
\end{enumerate}

\section{Conclusions}

\subsection{Take-home messages}

There are several messages that we would like to share with the reader:
\begin{itemize}
    \item There is an opportunity for a new computing paradigm where the hardware is stochastic by design.
    \item AI applications stand to benefit most from this hardware since many such applications are inherently stochastic.
    \item We identified a class of algorithms called Thermodynamic AI algorithms that not only use stochasticity as a resource but also employ a Maxwell's demon subroutine to guide the random fluctuations in the right direction. This provides a mathematical unification of seemingly unrelated algorithms, such as generative diffusion models, Bayesian deep learning, and Monte Carlo inference.
    \item Because of this unification, these algorithms can not only be run on a unified software platform, but can they can be run on a unified hardware paradigm. We identified the key ingredients of that hardware paradigm as stochastic-unit (s-unit) system coupled to a Maxwell's demon device.
    \item Varying degrees of physics can be incorporated into the Maxwell's demon device. We even consider the possibility of a fully physical Maxwell's demon based on forces and phase space dynamics. 
    \item Thermodynamic AI systems have some robustness to unintentional noise in the hardware. 
    \item Many of the tools of quantum computing, such as vector spaces, operators, and gate sequences, provide inspiration for analogous mathematics in this new paradigm of Thermodynamic AI.  
\end{itemize}

\subsection{Reaching distinct communities}

Regarding the last point, we would like to emphasize that this article is intended to reach a broad range of communities.

The first community of interest is the quantum computing (QC) one. The QC community is used to thinking of physics-based computing. Here we are pointing out a different opportunity for physics-based computing to the QC community. The other subtle message to the QC community is that noise can, in fact, be a resource. We could easily imagine taking everything said in this article and formulating an analogous version for quantum systems. Stochastic noise is ubiquitous in quantum systems and quantum Maxwell's demons are possible~\cite{lloyd1997quantum}. Hence, quantum thermodynamics could provide the physics-based paradigm for a novel form of quantum hardware that views stochastic fluctuations as a resource. 

The second community of interest is the AI one. AI algorithms like diffusion models are simultaneously breaking practical barriers and yet also not reaching their full potential due to a mismatch with the underlying hardware. Approximation-free Bayesian deep learning 
is currently prohibitively slow on digital hardware. Indeed, Thermodynamic AI hardware appears to be a natural paradigm to unlock major speedups for these AI algorithms.

On a final speculative note, there may be an opportunity to bridge the gap with communities studying intelligence in living systems, such as in neuroscience. Just as quantum computers are useful for simulating other quantum systems, there is the following question: Could Thermodynamic AI hardware provide a natural platform for simulating living intelligence?

\bibliography{thermo.bib}

\end{document}